\documentclass[reprint,aip,pof]{revtex4}

\usepackage{graphicx}
\usepackage{subfig}
\usepackage{multirow}
\graphicspath{{./fig/}}

\usepackage{bm}

\newcommand{\kf}{k_{\mathrm{f}}}

\begin{document}

\newcommand{\uu}{\mbox{\boldmath $u$} {}}
\newcommand{\FF}{\mbox{\boldmath $F$} {}}
\newcommand{\rr}{\mbox{\boldmath $r$} {}}
\newcommand{\vv}{\mbox{\boldmath $v$} {}}
\newcommand{\ww}{\mbox{\boldmath $w$} {}}

\newcommand{\EQ}{\begin{equation}}
\newcommand{\EN}{\end{equation}}
\newcommand{\EQA}{\begin{eqnarray}}
\newcommand{\ENA}{\end{eqnarray}}
\newcommand{\eq}[1]{(\ref{#1})}
\newcommand{\Eq}[1]{Eq.~(\ref{#1})}
\newcommand{\Eqs}[2]{Eqs.~(\ref{#1}) and~(\ref{#2})}
\newcommand{\eqs}[2]{(\ref{#1}) and~(\ref{#2})}
\newcommand{\Eqss}[2]{Eqs.~(\ref{#1})--(\ref{#2})}
\newcommand{\eqss}[2]{(\ref{#1})--(\ref{#2})}
\newcommand{\Section}[1]{Sec.\,\ref{#1}}
\newcommand{\Sec}[1]{\S\,\ref{#1}}
\newcommand{\App}[1]{Appendix~\ref{#1}}
\newcommand{\Fig}[1]{Fig.~\ref{#1}}
\newcommand{\Tab}[1]{Table~\ref{#1}}
\newcommand{\Figs}[2]{Figures~\ref{#1} and \ref{#2}}
\newcommand{\Tabs}[2]{Tables~\ref{#1} and \ref{#2}}
\newcommand{\bra}[1]{\langle #1\rangle}
\newcommand{\bbra}[1]{\left\langle #1\right\rangle}
\newcommand{\mean}[1]{\overline #1}
\newcommand{\meanEMF}{\overline{\mbox{\boldmath ${\cal E}$}} {}}
\newcommand{\meanFF}{\overline{\mbox{\boldmath ${\cal F}$}} {}}
\newcommand{\capback}{\eta_{\mathrm{back}}}
\newcommand{\meanB}{\overline{B}}
\newcommand{\meanF}{\overline{\cal F}}
\newcommand{\meanJ}{\overline{J}}
\newcommand{\meanU}{\overline{U}}
\newcommand{\meanT}{\overline{T}}
\newcommand{\meanrho}{\overline{\rho}}
\newcommand{\UU}{{\bm U}}
\newcommand{\UUp}{{\bm U}_{\rm p}}
\newcommand{\Lg}{L_{\rm g}}
\newcommand{\rhop}{\rho_{\rm p}}
\newcommand{\SSS}{{\sf S}}
\newcommand{\SSSS}{\mbox{\boldmath ${\sf S}$} {}}
\newcommand{\meanAA}{\overline{\mbox{\boldmath $A$}}}
\newcommand{\meanBB}{\overline{\mbox{\boldmath $B$}}}
\newcommand{\meanUU}{\overline{\mbox{\boldmath $U$}}}
\newcommand{\meanJJ}{\overline{\mbox{\boldmath $J$}}}
\newcommand{\meanEE}{\overline{\mbox{\boldmath $E$}}}
\newcommand{\meanuu}{\overline{\mbox{\boldmath $u$}}}
\newcommand{\meanAB}{\overline{\mbox{\boldmath $A\cdot B$}}}
\newcommand{\meanAoBo}{\overline{\mbox{\boldmath $A_0\cdot B_0$}}}
\newcommand{\ff}{\bm{f}}
\newcommand{\xx}{\bm{x}}
\newcommand{\kk}{\bm{k}}
\newcommand{\ii}{\mathrm{i}}
\newcommand{\eee}{\bm{e}}
\newcommand{\meanApoBpo}{\overline{\mbox{\boldmath $A'_0\cdot B'_0$}}}
\newcommand{\meanApBp}{\overline{\mbox{\boldmath $A'\cdot B'$}}}
\newcommand{\meanuxB}{\overline{\mbox{\boldmath $\delta u\times \delta B$}}}
\newcommand{\chk}[1]{[{\em check: #1}]}
\newcommand{\p}{\partial}
\newcommand{\xder}[1]{\frac{\partial #1}{\partial x}}
\newcommand{\yder}[1]{\frac{\partial #1}{\partial y}}
\newcommand{\zder}[1]{\frac{\partial #1}{\partial z}}
\newcommand{\xdertwo}[1]{\frac{\partial^2 #1}{\partial x^2}}
\newcommand{\xderj}[2]{\frac{\partial #1}{\partial x_{#2}}}
\newcommand{\timeder}[1]{\frac{\partial #1}{\partial t}}
\newcommand{\bec}[1]{\mbox{\boldmath $ #1$}}
\newcommand{\nab}{\mbox{\boldmath $\nabla$} {}}

\def\Rey{\mbox{\rm Re}}
\def\Nu{\mbox{\rm Nu}}
\def\Bi{\mbox{\rm Bi}}
\def\Pe{\mbox{\rm Pe}}
\def\Sc{\mbox{\rm Sc}}
\def\St{\mbox{\rm St}}

%
%
\newcommand{\yan}[3]{, Astron. Nachr. {\bf #2}, #3 (#1).}
\newcommand{\yact}[3]{, Acta Astron. {\bf #2}, #3 (#1).}
\newcommand{\yana}[3]{, Astron. Astrophys. {\bf #2}, #3 (#1).}
\newcommand{\yanas}[3]{, Astron. Astrophys. Suppl. {\bf #2}, #3 (#1).}
\newcommand{\yanal}[3]{, Astron. Astrophys. Lett. {\bf #2}, #3 (#1).}
\newcommand{\yass}[3]{, Astrophys. Spa. Sci. {\bf #2}, #3 (#1).}
\newcommand{\ysci}[3]{, Science {\bf #2}, #3 (#1).}
\newcommand{\ysph}[3]{, Solar Phys. {\bf #2}, #3 (#1).}
\newcommand{\yjetp}[3]{, Sov. Phys. JETP {\bf #2}, #3 (#1).}
\newcommand{\yspd}[3]{, Sov. Phys. Dokl. {\bf #2}, #3 (#1).}
\newcommand{\ysov}[3]{, Sov. Astron. {\bf #2}, #3 (#1).}
\newcommand{\ysovl}[3]{, Sov. Astron. Letters {\bf #2}, #3 (#1).}
\newcommand{\ymn}[3]{, Monthly Notices Roy. Astron. Soc. {\bf #2}, #3 (#1).}
\newcommand{\yqjras}[3]{, Quart. J. Roy. Astron. Soc. {\bf #2}, #3 (#1).}
\newcommand{\ynat}[3]{, Nature {\bf #2}, #3 (#1).}
\newcommand{\sjfm}[2]{, J. Fluid Mech., submitted (#1).}
\newcommand{\pjfm}[2]{, J. Fluid Mech., in press (#1).}
\newcommand{\yjfm}[3]{, J. Fluid Mech. {\bf #2}, #3 (#1).}
\newcommand{\ypepi}[3]{, Phys. Earth Planet. Int. {\bf #2}, #3 (#1).}
\newcommand{\ypr}[3]{, Phys.\ Rev.\ {\bf #2}, #3 (#1).}
\newcommand{\yprl}[3]{, Phys.\ Rev.\ Lett.\ {\bf #2}, #3 (#1).}
\newcommand{\yepl}[3]{, Europhys. Lett. {\bf #2}, #3 (#1).}
\newcommand{\pcsf}[2]{, Chaos, Solitons \& Fractals, in press (#1).}
\newcommand{\ycsf}[3]{, Chaos, Solitons \& Fractals{\bf #2}, #3 (#1).}
\newcommand{\yprs}[3]{, Proc. Roy. Soc. Lond. {\bf #2}, #3 (#1).}
\newcommand{\yptrs}[3]{, Phil. Trans. Roy. Soc. {\bf #2}, #3 (#1).}
\newcommand{\yjcp}[3]{, J. Comp. Phys. {\bf #2}, #3 (#1).}
\newcommand{\yjgr}[3]{, J. Geophys. Res. {\bf #2}, #3 (#1).}
\newcommand{\ygrl}[3]{, Geophys. Res. Lett. {\bf #2}, #3 (#1).}
\newcommand{\yobs}[3]{, Observatory {\bf #2}, #3 (#1).}
\newcommand{\yaj}[3]{, Astronom. J. {\bf #2}, #3 (#1).}
\newcommand{\yapj}[3]{, Astrophys. J. {\bf #2}, #3 (#1).}
\newcommand{\yapjs}[3]{, Astrophys. J. Suppl. {\bf #2}, #3 (#1).}
\newcommand{\yapjl}[3]{, Astrophys. J. {\bf #2}, #3 (#1).}
\newcommand{\ypp}[3]{, Phys. Plasmas {\bf #2}, #3 (#1).}
\newcommand{\ypasj}[3]{, Publ. Astron. Soc. Japan {\bf #2}, #3 (#1).}
\newcommand{\ypac}[3]{, Publ. Astron. Soc. Pacific {\bf #2}, #3 (#1).}
\newcommand{\yannr}[3]{, Ann. Rev. Astron. Astrophys. {\bf #2}, #3 (#1).}
\newcommand{\yanar}[3]{, Astron. Astrophys. Rev. {\bf #2}, #3 (#1).}
\newcommand{\yanf}[3]{, Ann. Rev. Fluid Dyn. {\bf #2}, #3 (#1).}
\newcommand{\ypf}[3]{, Phys. Fluids {\bf #2}, #3 (#1).}
\newcommand{\yphy}[3]{, Physica {\bf #2}, #3 (#1).}
\newcommand{\ygafd}[3]{, Geophys. Astrophys. Fluid Dynam. {\bf #2}, #3 (#1).}
\newcommand{\yzfa}[3]{, Zeitschr. f. Astrophys. {\bf #2}, #3 (#1).}
\newcommand{\yptp}[3]{, Progr. Theor. Phys. {\bf #2}, #3 (#1).}
\newcommand{\yjour}[4]{, #2 {\bf #3}, #4 (#1).}
\newcommand{\pjour}[3]{, #2, in press (#1).}
\newcommand{\sjour}[3]{, #2, submitted (#1).}
\newcommand{\yprep}[2]{, #2, preprint (#1).}
\newcommand{\pproc}[3]{, (ed. #2), #3 (#1) (to appear).}
\newcommand{\yproc}[4]{, (ed. #3), pp. #2. #4 (#1).}
\newcommand{\ybook}[3]{, {\em #2}. #3 (#1).}
\newcommand{\yjas}[3]{, J. Aerosol Sci. {\bf #2}, #3 (#1).}
\newcommand{\yast}[3]{, Aerosol Sci. Tech. {\bf #2}, #3 (#1).}
\newcommand{\yfue}[3]{, Fuel {\bf #2}, #3 (#1).}
\newcommand{\yate}[3]{, Atmospheric Env. {\bf #2}, #3 (#1).}
\newcommand{\ypps}[2]{, Part. Part. Syst. Charact., #2 (#1).}
\newcommand{\yjie}[2]{, J. Inst. Energy, #2 (#1).}

\title{The effect of turbulence on the particle impaction on a cylinder in a cross flow}
\author{Nikolai Hydle Rivedal}
\email{hydrivedal@gmail.com}
\affiliation{Department of Physics, The Norwegian University of Science and
Technology, H{\o}yskoleringen 5, N-7034 Trondheim, Norway}

\author{Anders Granskogen Bj{\o}rnstad}
\email{andersgb@gmail.com}
\affiliation{Department of Physics, The Norwegian University of Science and
Technology, H{\o}yskoleringen 5, N-7034 Trondheim, Norway}
\affiliation{Petrell, Trondheim, Norway}

\author{Nils Erland L. Haugen}
\email{Nils.E.Haugen@sintef.no}
\affiliation{SINTEF Energy Research, N-7034 Trondheim, Norway}

\date{\today}

\begin{abstract}
  Particle impaction on a cylinder in a cross flow is investigated 
  with the use of Direct Numerical Simulations (DNS) and with a
  focus on the effect of turbulence on the impaction efficiency. The
  turbulence considered is isotropic homogeneous turbulence with varying
  integral scales. It is found that for particles with Stokes numbers in
  the boundary stopping mode there
  is up to 10 times more front side impaction for turbulence with a large
  integral scale than for a corresponding laminar flow. 
  For decreasing integral scales the effect of
  the turbulence on front side particle impaction efficiency is decreasing. 
  The back side impaction efficiency is also found to be 
  influenced by the turbulence, but for the back side the strongest effect,
  and largest impaction efficiency, is found for small integral scales.
\end{abstract}
\pacs{
}
\maketitle

\section{Introduction}
Particle-laden fluid flows are common both in nature and in 
a large number of industrial applications. 
Depending on the particle and fluid properties, particles may impact on
a solid object entrained in the fluid flow. Such particle impaction may either
lead to the buildup of a deposition layer on the solid-fluid interface,
or to erosion of the solid object. Typical industrial
applications are
filters, furnaces, industrial boilers, fan cooled 
electronics and ventilation systems.

For several of the applications mentioned above the solid object may 
be approximated by a cylinder. Due to this, and to the simplicity of the
cylindrical geometry, determination of particle impaction 
efficiency on a cylinder in a cross flow has become a benchmark case.
By using potential flow theory for the fluid flow 
particle impaction efficiencies can been found
\cite{israel1983,baxter1990,ingham1990} 
as a function of the particle size. Since potential flow
theory assumes infinite Reynolds numbers, but still does not account 
for turbulence,
it is clear that this method has major shortcomings. It is nevertheless
a well accepted method for Stokes numbers larger than unity. 
The Stokes number 
$\St=\tau_p/\tau_f$ 
is the ratio of the particle Stokes time
to the timescale of the fluid flow around the solid object.
Numerous other approaches focusing on experimental 
\cite{schweers1994,kasper2009},
numerical CFD
\cite{suneja1974,yilmaz2000,li2008}
and phenomenological modelling
\cite{loehden1989,huang1996}
are also found in the literature. 

As the Reynolds number is increased turbulence will become 
important and for very large Reynolds numbers turbulence will significantly
alter the result. 
Turbulence is either generated due to strong shear induced by the cylinder, 
or, alternatively, the flow approaching the
cylinder is already turbulent. For $\Rey=U_0D/\nu \sim 10^3 $ the wake downstream
of the cylinder will break up into turbulence even when the flow is laminar 
upstream of the cylinder. 
For such low 
Reynolds numbers turbulence is not expected to have any impact on the 
particle impaction efficiency as the transition to turbulence appears too
far downstream of the cylinder. Here $U_0$ is the mean 
flow velocity, $D$ is the
cylinder diameter and $\nu$ is the kinematic viscosity.  As the Reynolds number
increases, the transition to turbulence moves upstream towards the cylinder
and for $\Rey\sim 10^4$, the boundary layer around the cylinder will be
turbulent. 

The effect of turbulence on particle impaction has been 
studied by Douglas \& Ilias~(1988)~\cite{douglas1988}. In this study the cylinder was situated within
a channel such that the turbulence was generated by the turbulent channel
flow and not by the cylinder itself. 
Except from this, not much is found in the literature
on the effect of turbulence on the impaction efficiency on a cylinder in
a cross flow. There is however a large number of publications on particle
impaction on the walls of a turbulent channel flow, 
where the particles are impacting on
the channel walls due to turbophoresis (see e.g. Picano et al. \cite{picano2009} 
and references therein). For a cylinder in a cross flow, turbophoresis is 
expected to be relevant for very large Reynolds numbers
$\Rey > 10^4$.

\section{The effect of turbulence on impaction efficiency}
As already mentioned, turbulence can be introduced in two 
different ways; either the turbulence is generated due to the shear
induced by the cylinder, or turbulence is generated somewhere upstream
of the cylinder and is inherent in the flow as it approaches the cylinder.
The current work focuses only on turbulence generated upstream of the cylinder,
and does not consider Reynolds numbers large enough for turbulence to be
generated in the boundary layer around the cylinder. Furthermore it is 
assumed that the turbulence approaching the cylinder is homogeneous and
isotropic.

If the turbulence is generated upstream of the cylinder there are
several possible scenarios, some of which will be explained
in the following.

\subsection{Large eddy Stokes numbers}
The eddy turnover time is given as
\EQ
\tau_e=d/u_{\rm rms},
\EN
where $d$ is the integral scale of the turbulence and 
$u_{\rm rms}=\sqrt{\langle(U-U_0)^2\rangle}$ 
is 
the root mean square velocity of the turbulence and $\langle\rangle$ 
symbolizes ensemble averaging. 
If the eddy turnover time is much shorter than 
the particle Stokes time, i.e. $\St_{\rm eddy}=\tau_p/\tau_e >>1 $, 
the particles are too slow
to be affected by the turbulent eddies. When this is the case the turbulence
has no effect on the particles, and consequently the impaction efficiency 
is not affected. Here the particle Stokes time is 
\EQ
\tau_p=\frac{Sd_p^2}{18\nu},
\EN
where $S=\rho_p/\rho$, $\rho$ is the fluid density, $\rho_p$ is the 
particle material density and $d_p$ is the particle diameter.

\subsection{Large scale turbulence}
It is well known that the impaction efficiency $\eta=N_{\rm impact}/N_{\rm init}$ 
in a laminar flow is determined
by the Stokes number \cite{haugen_kragset10}. Here the number of particles
impacting on the cylinder surface is given by $N_{\rm impact}$ while 
$N_{\rm init}$ represents the number of particles initially positioned such that 
the mean flow at their position is
moving in the direction of the cylinder.
When the eddy turnover time is equal to or larger than the particle Stokes time,
i.e. $\tau_e \ge \tau_p$,
the particles will follow the turbulent eddies. Consequently, the particle velocities will deviate from the mean flow velocity. 
When the scale of the turbulence is not small compared to the size of the 
cylinder this yields a modified 
Stokes number, which will then also give a change in the particle
impaction efficiency.

The Stokes number $\St=\tau_p/\tau_f$ is proportional to the mean fluid flow 
velocity since $\tau_f=D/U_0$. 
In the turbulent case the magnitude of the fluid flow velocity 
$U$, in general different from the mean flow velocity $U_0$, 
is stochastic. 
When the integral scale of the turbulence is large, i.e. $d\gtrsim D$, $U$ may
be seen as the mean flow velocity at that instant in time.
Thus, the Stokes number also is a stochastic variable, 
effectively being different from the 
'laminar' $\St$, expressed by the laminar fluid velocity $U_0$. 
The effective Stokes number can be expressed as
\begin{equation}
\label{eq:turbStokes}
\St_{\mathrm{eff}}=\frac{\St}{U_0}U.
\end{equation} 
As $\St_{\mathrm{eff}}$ is a linear function of $U$, its variance becomes   
\begin{equation}
\label{eq:varianceSt}
Var(\St_{\mathrm{eff}})\equiv \sigma^2_{\St}=\bigg{(}\frac{\St}{U_0}
\bigg{)}^2Var(U),
\end{equation}
cf. Walpole et al. (2007) \cite{statistics}. 
Since 
$U_0$ is constant, (\ref{eq:varianceSt}) shows 
that $\sigma^2_{\mathrm{St}}=0$ when $U=U_0$. The expectation value 
of the Stokes number equals 
\begin{equation}
\label{eq:expectStokes}
E(\St_{\mathrm{eff}})=\St,
\end{equation}
since $U$ fluctuates symmetrically 
around $U_0$. 

With a fluctuation in $U$, the 
effective Stokes 
number becomes $\St_{\mathrm{eff}}=\St+\Delta$, 
with $\Delta=\St \delta_U/U_0$ being 
the resulting fluctuation in the Stokes number when $\delta_U=U-U_0$.
Thus, 
a Taylor expansion in the small parameter $\Delta$ yields, 
by using (\ref{eq:expectStokes}),
\begin{eqnarray}
\eta(\St_{\mathrm{eff}})&=&\eta(\St+\Delta)\nonumber\\
&=&\eta(\St)+\eta'(\St)\Delta+\frac{\eta''(\St)}{2}
\Delta^2+\mathcal{O}(\Delta^3).
\end{eqnarray}
The expectation value of this becomes
\begin{eqnarray}
\label{eq:expectedCapEff}
E[\eta(\St_{\mathrm{eff}})]&=&\eta(\St)+\eta'(\St)E[\Delta]+
\frac{\eta''(\St)}{2}E[\Delta^2]+\mathcal{O}(E[\Delta^3])\nonumber
\\
&\approx& \eta(\St)+\frac{\eta''(\St)}{2}\sigma^2_{\St}.
\end{eqnarray}
By definition, $\sigma^2_{\mathrm{\St}}\equiv E[\Delta^2]-
(E[\Delta])^2=E[\Delta^2]$ since $E[\Delta]=0$ 
due to the symmetry of the velocity fluctuations around the mean.  
Furthermore, higher order terms, $\mathcal{O}(E[\Delta^3])$, have 
been neglected. 
When there is no turbulence, such that $\Delta =0$, the expectation value
of the impaction efficiency is 
$E(\eta(\St_{\rm eff}))=\eta(\St)$, i.e. it equals the impaction efficiency in
the laminar case, as expected. 

In order 
to use \Eq{eq:expectedCapEff} to obtain the expected values of 
the front side impaction efficiencies with turbulence present, values for 
$\eta''(\St)$ are required.

Furthermore, $\sigma^2_{\St}$ is needed to determine the expectation value
of the front side 
impaction efficiency $\eta(\St_{\mathrm{eff}})$ for a given $\St$. 
For non-inertial particles, i.e. $\St\rightarrow 0$, particles follow the fluid 
flow exactly and  $\sigma^2_{\St}=u_{\rm rms}$.
For non-zero Stokes numbers the particles do not follow the fluid flow
and $\sigma^2_{\St}$
must be found from the simulations. This is done by looking at how 
the velocity of the particles deviate from the mean flow velocity 
$U_0$. This deviation expresses how the turbulence affects the 
particles, and will typically be different for different Stokes 
numbers. 
To find reliable values of  $\sigma^2_{\St}$, the position where
$\sigma^2_{\St}$ is measured,
should not be too close to 
the cylinder.

%

\subsection{Small scale turbulence}
If the turbulent eddies are very small, they may even penetrate 
into the boundary layer around the cylinder. If this happens, the particles can
impact on the cylinder surface due to turbophoresis, which may have a
significant effect on the impaction efficiency.
Let us call this effect impaction by external turbophoresis. 
The boundary layer thickness around the cylinder
is given by \cite{haugen_kragset10}
\EQ
\delta=\frac{BD}{\sqrt{\Rey}}
\EN
where $B\approx 0.5$ such that in order to have impaction by external 
turbophoresis the turbulent integral scale $d$ must fulfill the inequality
\EQ
\label{dless}
d<\frac{D}{2\sqrt{\Rey}}.
\EN
The dissipative term, in spectral space, is given as $\nu k^2 (U-U_0)$ such that
for small scale turbulence, i.e. when the wavenumber $k$ 
is large, the decay will be very fast.
Consequently, if the turbulence generated upstream of the cylinder 
shall survive all the 
way down to the cylinder, the timescale of the eddy decay, given by
$\tau_\nu=d^2/\nu$, must be longer than the typical fluid timescale
$\tau_f=D/U$ such that
\EQ
\label{visc}
\Rey>\frac{1}{4}\left(\frac{D}{d}\right)^2.
\EN
By combining \Eq{dless} and \Eq{visc} it is found that
\EQ
\Rey>\frac{1}{4}\left(\frac{D}{d}\right)^2>4\Rey.
\EN
It is clear that the above inequality is a contradiction, and consequently 
the conclusion can be drawn that impaction by external turbophoresis, where the
scale of the turbulence is small enough to penetrate the boundary layer around
the cylinder,
is not practically feasible if the source of turbulence is not very close
to the cylinder. An example where the source is indeed close to the
cylinder is a cylinder placed in a turbulent channel flow \cite{douglas1988}. 
This is a situation which is encountered in many industrial applications, 
but it does introduce some extra parameters into the study.
Due to the increased parameter space this application is not considered here; 
instead, this work focuses solely on isotropic decaying turbulence.

\section{Simulations}
The DNS code used for the simulations performed in this work is 
\textsc{The Pencil Code} \cite{PC}. 
The fluid flow is evolved in time by solving the continuity equation
\EQ
\frac{\partial \rho}{\partial t}+\nabla\cdot(\uu\rho)=0
\EN
and the momentum equation
\EQ
\label{mom}
\frac{\partial \uu}{\partial t}+\uu\cdot\nabla \uu+\frac{1}{\rho}\nabla P
=\frac{1}{\rho}\nabla\cdot\left(2\rho\nu \SSSS\right)+\frac{\FF}{\rho}
\EN
together with the equation of state given by $P=\rho c_s^2$. Here 
$\uu$ is the fluid velocity,
$c_s$ is the speed of sound,
$P$ is the pressure,
$\FF$ is some external force and
$\SSSS$ is the traceless rate of strain tensor.

The particles are tracked individually in a Lagrangian manner as
\EQ
\frac{d\vv}{dt}=\FF_p
\EN
and
\EQ
\frac{d\xx}{dt}=\vv,
\EN
when $\vv$ is the particle velocity, 
$\xx$ is the particle position and
$\FF_p=(\uu-\vv)/\tau_p$ is the drag force from the fluid on the particles.

In the following, all variables are non-dimensionalized by the 
mean flow velocity $\tilde{U}_0$, and
the length of one of the sides of the simulation box $\tilde{L}_{\rm box}$.

Isotropic turbulence is created
inside a cubic domain, the \emph{turbulent box}. The length of each of the sides
of the box is $L_{\rm box}=1$ and two different Reynolds
numbers are simulated, namely 420 and 1685.  
At $\Rey=420$, the number of grid points are $512^3$, while at $\Rey=1685$,
 $1024^3$ grid points are required.   
Statistically stationary turbulence is
achieved by the use of external forcing. 
Energy is added at a given wave number, and the turbulence develops until 
the energy input through the forcing is equal to the energy dissipated in 
the dissipative subrange. This leads to an equilibrium 
where fluid motion is independent of the turbulence's 
initial conditions \cite{boivin} and the turbulent flow
is statistically stationary. 
The forced turbulence is created by adding a 
stochastic forcing function
\begin{equation}
\label{forcing}
\FF(\boldsymbol{x},t)=\Re \lbrace N\mathbf{f}_{\boldsymbol{k}(t)}\exp[\mathit{i}\boldsymbol{k}(t)\cdot\boldsymbol{x}+\mathit{i}\phi(t)]\rbrace
\end{equation}
as the external force on the right hand side of the momentum equation (\Eq{mom}). 
As $\boldsymbol{k}(t)$ and $\phi (t)$ are chosen 
randomly at each time step, the stochastic nature of turbulence 
is inherent in the equation. The normalization factor is 
$N=f_{0}c(|\boldsymbol{k}|c/\Delta t)^{1/2}$ and $\mathbf{f}_{\boldsymbol{k}(t)}$ is perpendicular to $\boldsymbol{k}$ and
an eigenfunction to the curl operator.  
A forcing wavenumber  
$k_{\mathrm{f}}$ is chosen such that
the magnitude of the randomly chosen $\boldsymbol{k}(t)$ 
is in the range 
$k_{\mathrm{f}}-0.5 < |\boldsymbol{k}(t)|/k_0 < k_{\mathrm{f}}+0.5$, where $k_0=1/L_{\rm box}$ is the normalizing 
wavenumber corresponding to the side lengths $L_{\rm box}$ 
of the box. 
For more details on the forcing see e.g.
Haugen \& Brandenburg (2006)~\cite{HB06}.
The effect of the forcing is that turbulent energy 
is put into the system at the spatial scale corresponding to 
the forcing wavenumber $\kf$, which in this manner 
determines the behaviour of the turbulence. As $\kf$ 
is normalized by $k_0$, the forcing length scale will be 
$d=L_{\rm box}/\kf=\kf^{-1}$. 

\begin{figure}[h!]
  \subfloat[Turbulent box.]{%
    \label{fig:first}
    \includegraphics[width=0.4\textwidth]{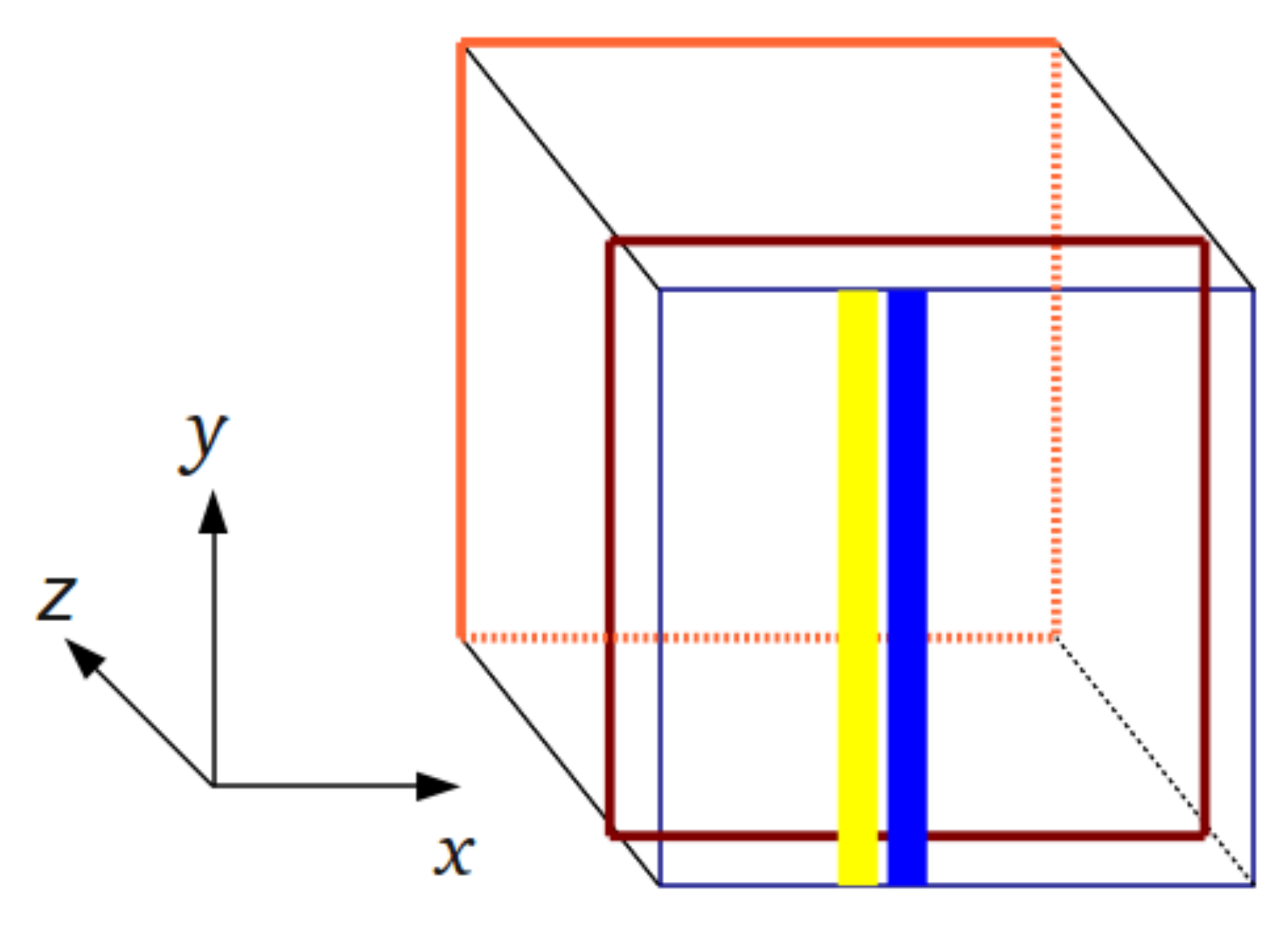}
  }\\
  \subfloat[Turbulent inlet on the domain.]{%
    \label{fig:second}
    \includegraphics[width=0.4\textwidth]{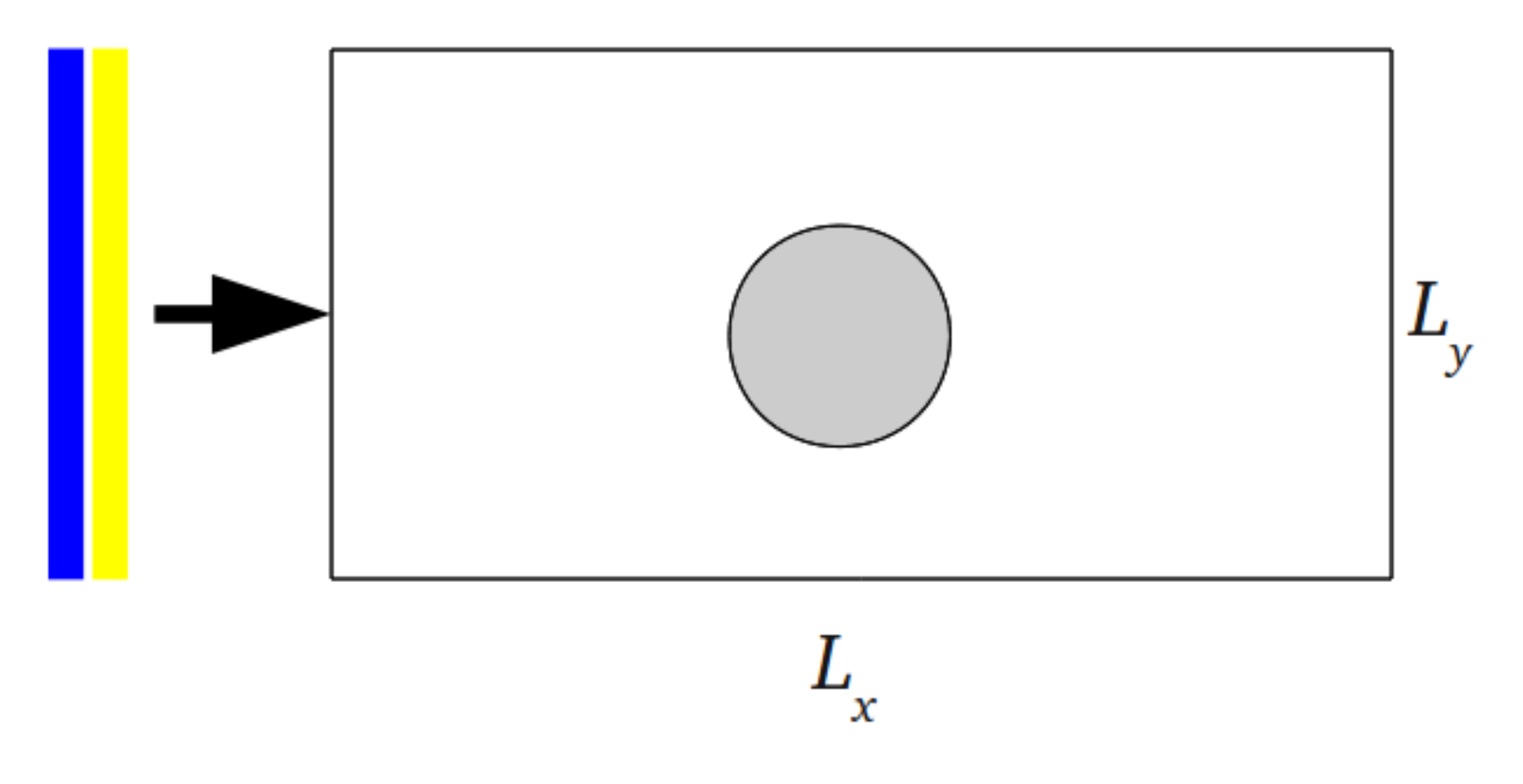}
  }
  \caption{Figure (a) shows how turbulent velocity 
    information containing strips (yellow and blue here) 
    at discrete increasing $x$-values are successively extracted 
    from the $xy$-plane, i.e. for constant $z$. As all strips 
    for $z=0$ are extracted, in the blue marked plane, the 
    procedure is repeated at the next discrete $z$-value. When the 
    last $xy$-plane, the orange one, is used, the procedure starts 
    over again. In (b) it is shown how the turbulent velocities of 
    the successive strips are imposed on the boundary of the domain.       
  }
    \label{NSCBC}
\end{figure}       
The turbulence is imposed on the inlet of the 
two-dimensional main domain (see \Fig{fig:second}) 
of size $L_x\times L_y=2\times 1$. 
The cylinder cross section with a diameter $D=L_{\rm box}/6$ 
is placed in the center of the 2D domain.
The fluid flow enters the domain with a mean velocity $U_0=1$. 
Prior to letting the turbulence enter the domain, 
von K\'{a}rm\'{a}n vortices in a steady state are established 
in the wake of the cylinder. When this is done, the 
turbulence, generated in the box shown in \Fig{fig:first}, 
is used as inlet for the two dimensional 
domain containing the flow, the cylinder and the particles. 
As illustrated in \Fig{NSCBC}, 
a quadratic $xy$-slice of the turbulent 
box is divided into strips whose velocity information is 
extracted and inserted at the leftmost side of the domain,
shown in \Fig{fig:second}. 
At successive time steps, the position of the strips chosen 
depends on the distance traveled by the fluid, i.e. 
$U_{0}t$. When the end of a slice is used as inlet, the slice 
at the next discrete $z$-value is used in the same manner. 
When the outer end of the box is used, i.e. the slice at 
the maximum $z$-value, the procedure is started over again 
with the strips from the slice at $z=0$. Imposing turbulence in 
this way essentially means adding the turbulent velocity 
$\boldsymbol{u}_{\mathrm{t}}$ to the velocity 
$U_0\boldsymbol{\hat{x}}$. Hence, the velocity 
at the inlet is
\begin{equation}
\label{equ:turbInlet}
\boldsymbol{u}_{\mathrm{in}}=U_0\boldsymbol
{\hat{x}}+\boldsymbol{u}_{\mathrm{t}}.
\end{equation}
The boundary conditions implemented when imposing the 
turbulent velocity at $x=0$ are the \emph{Navier-Stokes characteristic boundary conditions} (NSCBC) \cite{poinsot_lele}. The advantage 
of using NSCBC is that the boundaries are non-reflecting, 
meaning that any signal is let through them. The direct 
numerical simulation of compressible flows requires an 
accurate control of wave reflections from the domain 
boundaries, as the accuracy of the solution is in general 
sensitive to solutions at the boundaries.
In compressible fluid flows, and especially in DNS where 
the range of scales is large, 
reflected waves can cause problems. This is due to e.g. convected 
vorticity or sound waves not being let through the boundaries, but rather
being reflected back into the domain. 
This may even lead to strong standing waves in the computational domain.
The transparency of NSCBC prevents this. 
In the current work the NSCBC method as described by Lodato 
et al. (2008) \cite{lodato08} is used
both for the inlet and the outlet.

The 2D domain containing the cylinder cross section and 
the surrounding flow, where the flow is in the x-direction, 
has periodic boundary conditions 
in the $y$-direction, meaning that a particle or a 
fluid element hitting the boundary at $y=0$ or $y=L_{y}$ 
immediately is reinserted at the same $x$-position on 
the other side, with the same velocity and in the same state. 

Particles are inserted at $x=0$, and are removed from the 
simulation when hitting the cylinder or the rightmost boundary at $x=L_{x}$. 
Among similar simulations done in the past, 
there have been some disagreements related to the correct 
number of particles to insert to achieve the desired 
statistical reliability of the data \cite{strutt}. To ensure 
statistically significant data, a large number of 
particles, $N=10^6$, is inserted in each simulation. The 
particles are inserted with an initial velocity $V_0=U_0$
at a rate of $0.133\cdot 10^6$ particles per time unit, where the time unit
has been non-dimensionalized by $\tilde{t}=\tilde{L}_{\rm box}/\tilde{U}_0$.
This correspond to a total particle insertion time of
several von K\'{a}rm\'{a}n eddy times, which is required in order to get
statistically stationary results.  
An overview of the 
two-dimensional simulations is given in table \ref{tab:cases}.
\begin{table}[h!]
\caption{Overview of the 2D simulations. 'L' corresponds to lower $\Rey$ runs and 'H' to higher.}
\label{tab:cases}
\begin{center}
\begin{tabular}{c|c|c|c}
Case & Resolution & $\Rey$ & Flow regime\\
\hline
L1 & $1024 \times 512$ & 420 & Laminar\\
L2 & $1024 \times 512$ & 420 & Turbulent; $\kf=15$\\
L3 & $1024 \times 512$ & 420 & Turbulent; $\kf=5$\\
L4 & $1024 \times 512$ & 420 & Turbulent; $\kf=1.5$\\
\hline \hline
H1 & $2048 \times 1024$ & 1685 & Laminar\\
H2 & $2048 \times 1024$ & 1685 & Turbulent; $\kf=15$\\
H3 & $2048 \times 1024$ & 1685 & Turbulent; $\kf=5$\\
H4 & $2048 \times 1024$ & 1685 & Turbulent; $\kf=1.5$\\
\hline
\end{tabular}
\end{center}
\end{table}

\section{Results}
\begin{figure}
  \centering
  \includegraphics[width=0.5\textwidth]{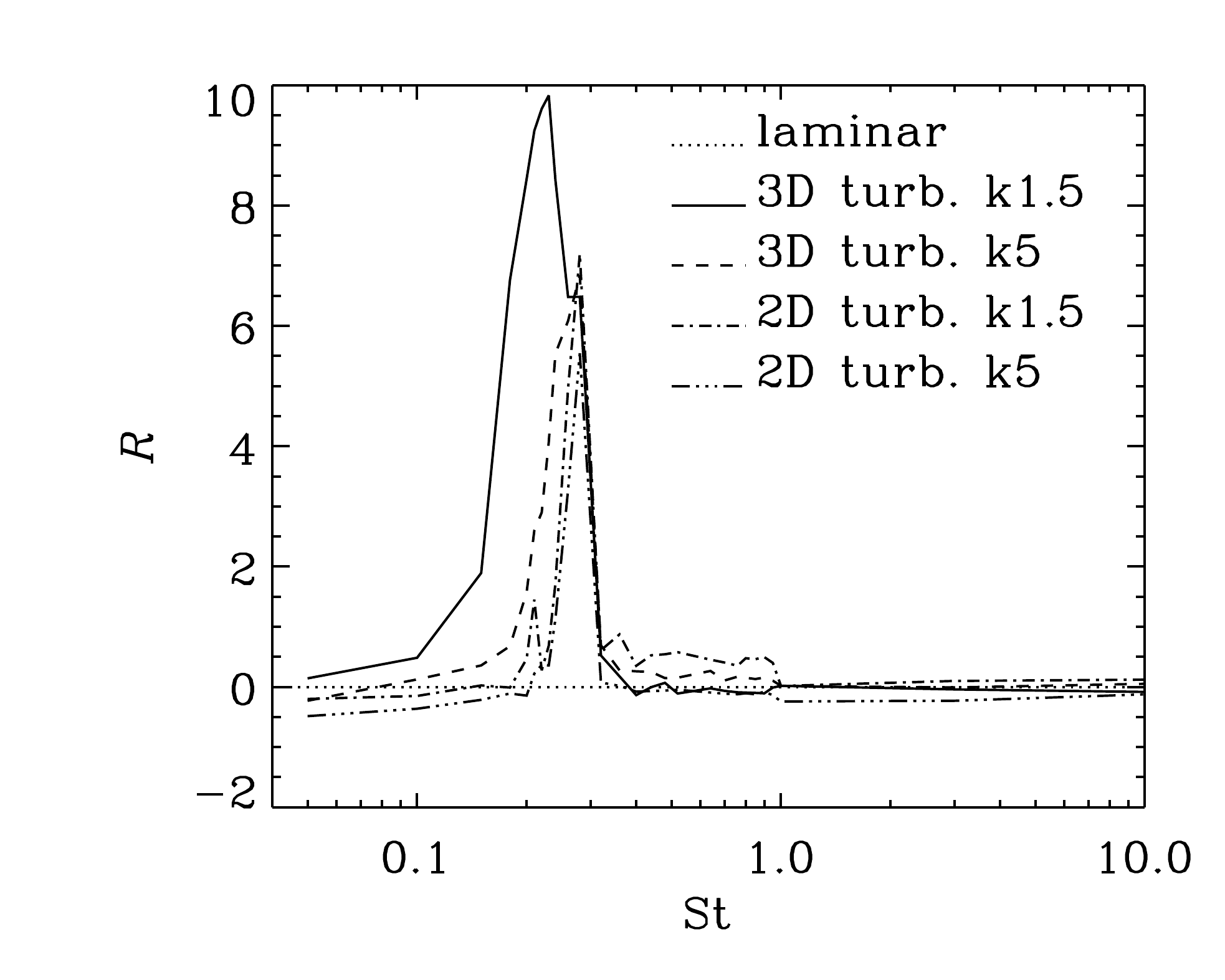}
  \caption{Relative differences in impaction efficiency for two and three 
dimensional simulations with forcing at $k_f=1.5$  and $k_f=5$ and $\Rey=420$.}
  \label{fig:2D_reldiff}
\end{figure}

The relative difference in impaction
efficiency between a turbulent and a laminar simulation is defined as
\EQ
R=\frac{\eta -\eta_{\rm lam}}{\eta_{\rm lam}}
\EN
where $\eta$ is the turbulent impaction efficiency and
$\eta_{\rm lam}$ is the impaction efficiency of the corresponding non-turbulent
setup.
In \Fig{fig:2D_reldiff} the relative impaction efficiency is shown for
simulations with $\Rey=420$ and for forcing at $k_f=1.5$ and $k_f=5$ both
for two and three dimensional simulations. 
It is clear that there are some
differences between the two and the three dimensional results, but qualitatively
the results are similar. In the following the focus will therefore be on the
two dimensional results as three dimensional simulations on Reynolds numbers
larger than 420 can not be afforded.

\subsection{Effect of turbulent integral scale for $\Rey=1685$}
\begin{figure}[h!]
\centering
\includegraphics[width=0.5\textwidth]{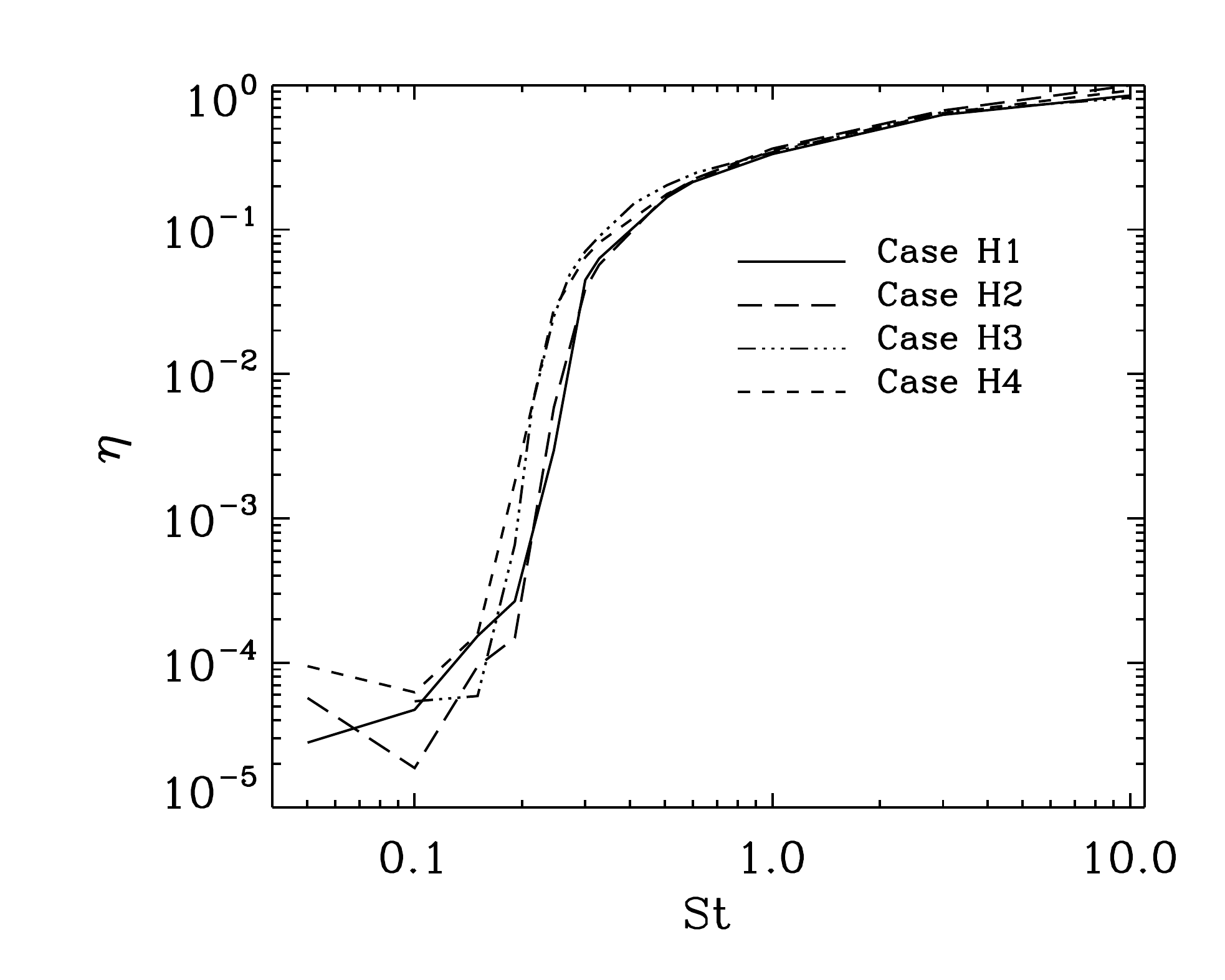}
\caption{Front side impaction efficiencies at $\Rey$=1685,
 for the laminar reference case and the three turbulent forced
simulations.}
\label{fig:Re1685}
\end{figure}

\begin{figure}
\centering
\includegraphics[width=0.5\textwidth]{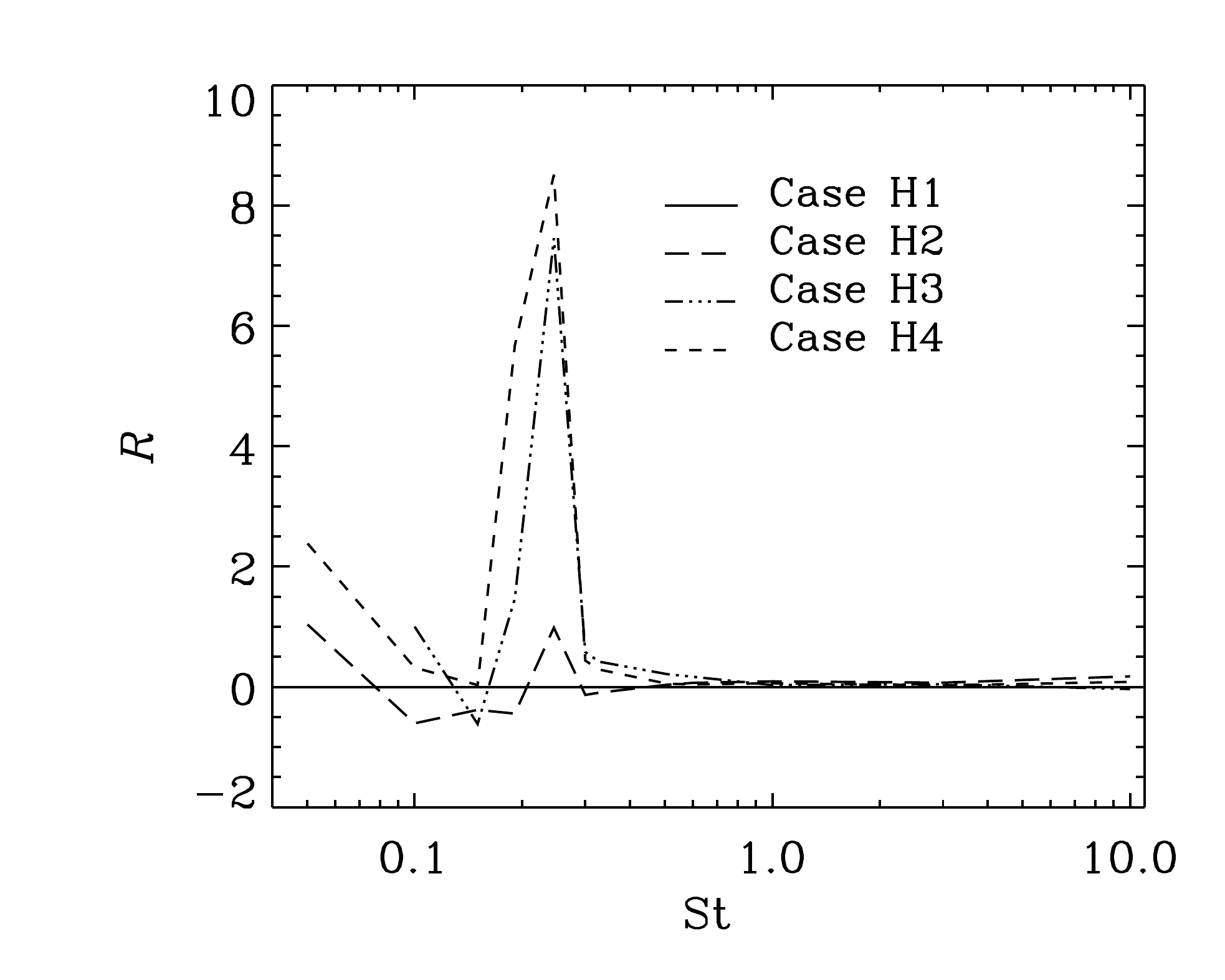}
\caption{Front side impaction efficiencies for the cases with turbulence at 
$\Rey$=1685, plotted as relative difference to the laminar.}
\label{fig:Re1685_reldiff}
\end{figure}

Figure \ref{fig:Re1685} shows the front side impaction efficiency plots,
with their characteristic shape, for simulations with $\Rey=1685$. 
The boundary stopping mode is
in the range $0.15\lesssim\St\lesssim 0.5$, with the
inertial impaction and boundary interception modes above and below
this, respectively. 
The effects of turbulence at $\Rey=1685$, by
means of the different forcing wave numbers $\kf$, are investigated by
examining the plotted relative differences in front side impaction
efficiency in \Fig{fig:Re1685_reldiff}. 
The peak in relative difference
is clear for all three cases, and is found at $\St=0.24$. So the effect
of turbulence is largest in the lower region of the boundary
stopping mode. The increase in the front side impaction efficiency is as
expected stronger for the larger forcing length scales.
For $\St=0.24$ the impaction efficiency is almost 10 times higher 
for a turbulent flow with an integral scale at $k_f=1.5$ 
than for a corresponding laminar flow.
The case with $\kf=5$ also
has a dramatic increase in impaction, compared to the laminar, from
$\St\approx 0.05$ on, while the change is much less for the $\kf=15$
case. It should also be noted that the same effects of the turbulence are
found at the same Stokes numbers for all forcing scales.

\subsection{Comparison with predictions for large scale turbulence}
\begin{figure}
\centering
\includegraphics[width=0.5\textwidth]{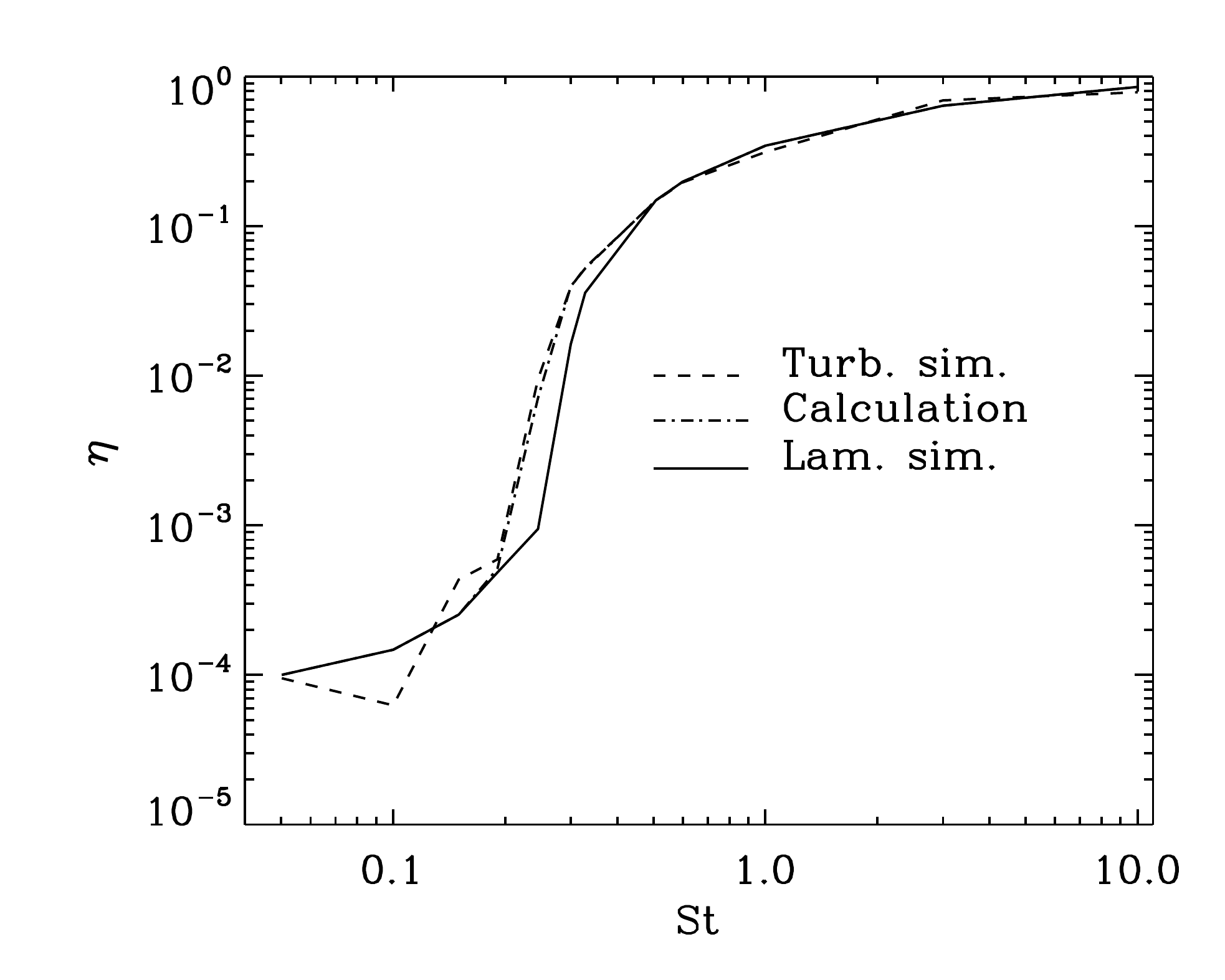}
\caption{Calculated impaction efficiency with $\kf=1.5$ turbulence 
at $\Rey=420$, plotted together with the impaction efficiencies 
of case L4 (Turbulent simulation) and case L1 (Laminar simulation).}
\label{fig:calcEta_k1.5_421}
\end{figure}

\begin{figure}
\centering
\includegraphics[width=0.5\textwidth]{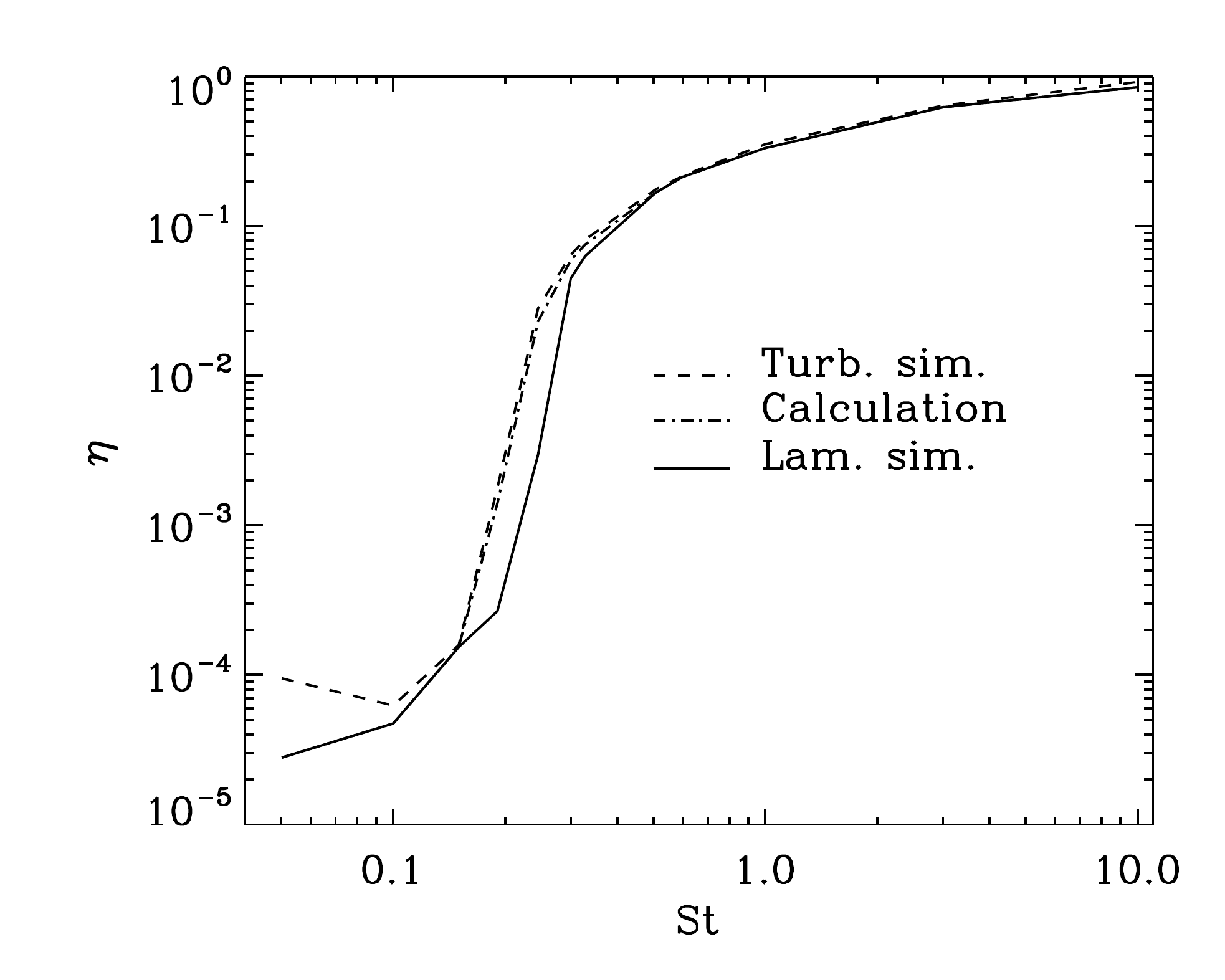}
\caption{Calculated impaction efficiency with $\kf=1.5$ turbulence 
at $\Rey=1685$, plotted together with the impaction efficiencies of 
case H4 (Turbulent simulation) and case H1 (Laminar simulation).}
\label{fig:calcEta_k1.5_1685}
\end{figure}

\begin{figure}[h!]
\centering
\includegraphics[width=0.5\textwidth]{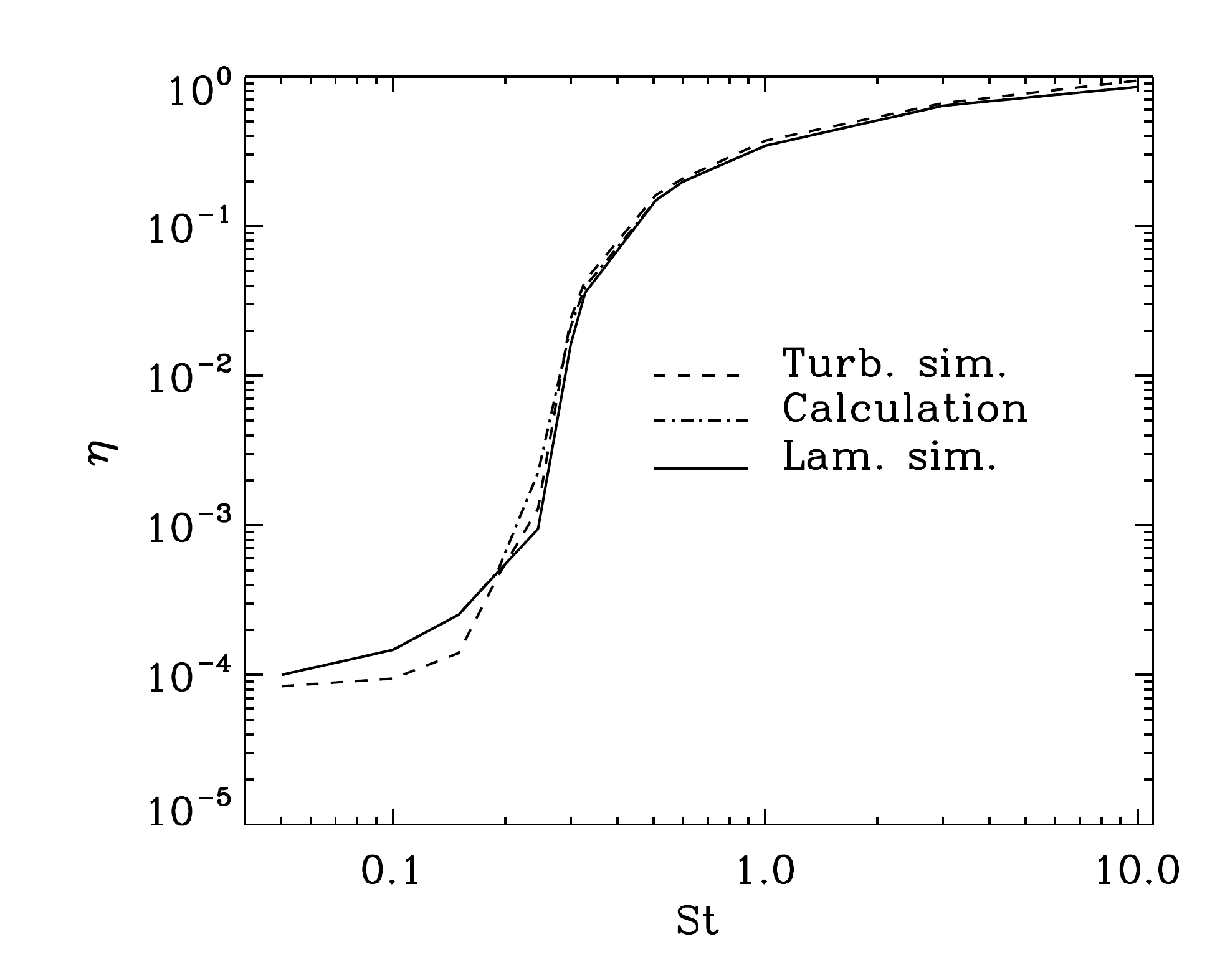}
\caption{Calculated impaction efficiency with $\kf=5$ turbulence 
at $\Rey=420$, plotted together with the impaction efficiencies of 
case L3 (Turbulent simulation) and case L1 (Laminar simulation).}
\label{fig:calcEta_k5_421}
\end{figure}

\begin{figure}[h!]
\centering
\includegraphics[width=0.5\textwidth]{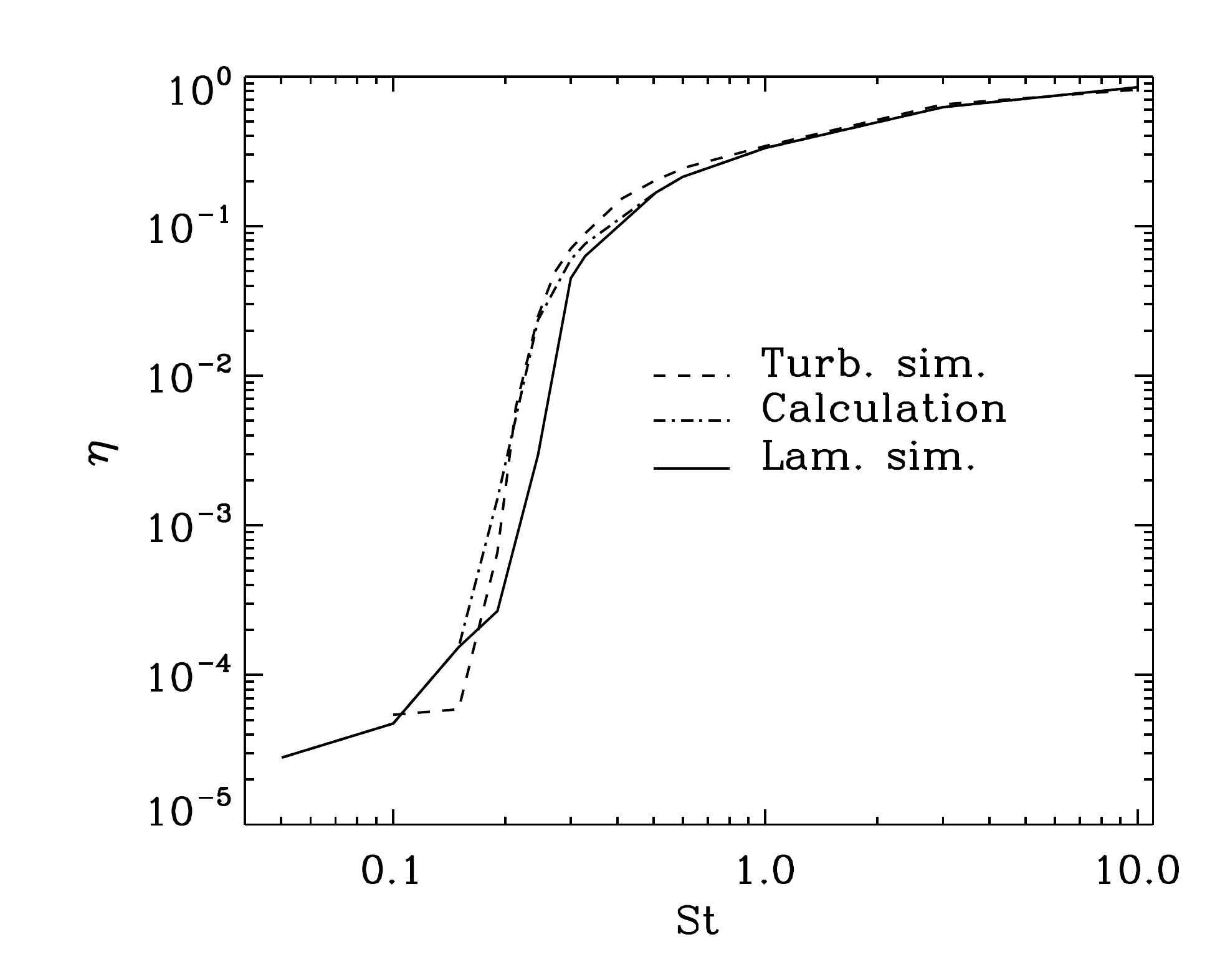}
\caption{Calculated impaction efficiency with $\kf=5$ turbulence 
at $\Rey=1685$, plotted together with the impaction efficiencies 
of case H3 (Turbulent simulation) and case H1 (Laminar simulation).
For $\St<0.1$, the capture efficiency in the turbulent simulation
is zero.}
\label{fig:calcEta_k5_1685}
\end{figure}

\begin{figure}[h!]
\centering
\includegraphics[width=0.5\textwidth]{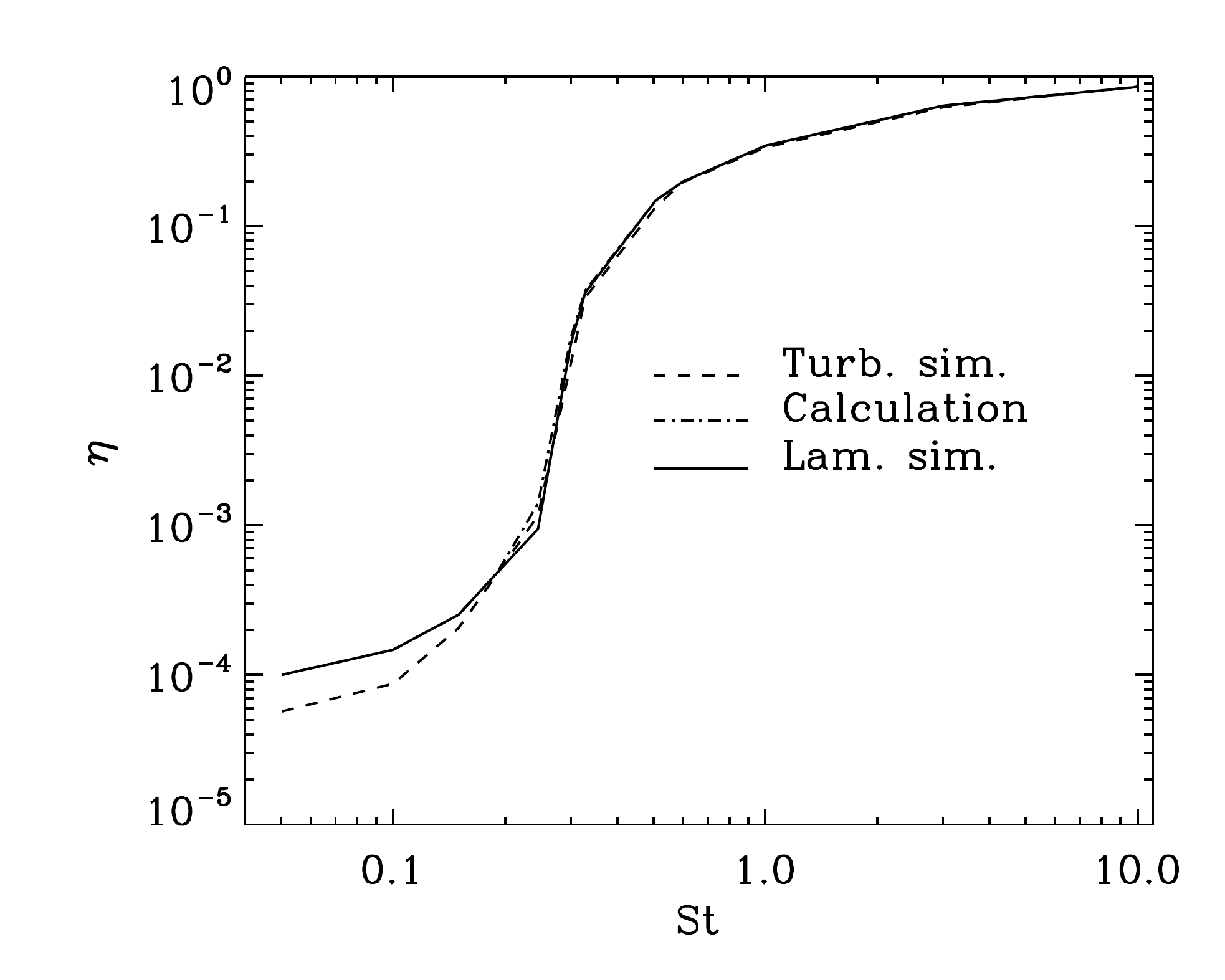}
\caption{Calculated impaction efficiency with $\kf=15$ turbulence 
at $\Rey=420$, plotted together with the impaction efficiencies
 of case L2 (Turbulent simulation) and case L1 (Laminar simulation).}
\label{fig:calcEta_k15_421}
\end{figure}

\begin{figure}[h!]
\centering
\includegraphics[width=0.5\textwidth]{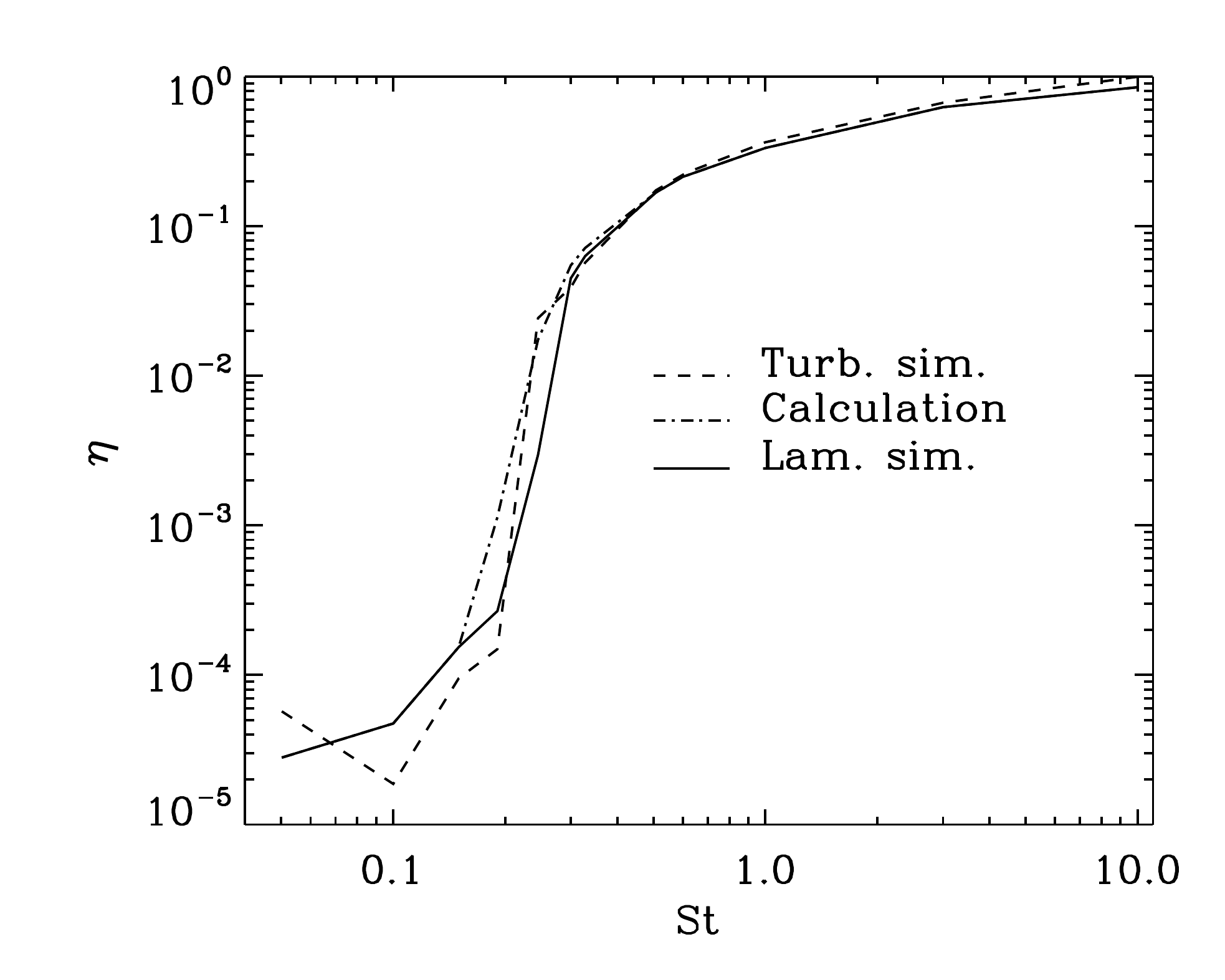}
\caption{Calculated impaction efficiency with $\kf=15$ turbulence
 at $\Rey=1685$, plotted together with the impaction efficiencies
 of case H2 (Turbulent simulation) and case H1 (Laminar simulation).}
\label{fig:calcEta_k15_1685}
\end{figure}

In this subsection numerical results will be compared with results obtained
from \Eq{eq:expectedCapEff}.
The expected impaction efficiencies, calculated according to 
\Eq{eq:expectedCapEff}, are plotted in Fig. 
\ref{fig:calcEta_k1.5_421} through Fig. \ref{fig:calcEta_k15_1685}.
The results from the simulations at $\kf=1.5$ (\Figs{fig:calcEta_k1.5_421}{fig:calcEta_k1.5_1685}) 
are found to be as predicted by the calculations, in 
the boundary stopping and the classical impaction modes. 
However, in the boundary interception mode for $\St\lesssim 0.15$,
the results from the simulations are seen to deviate from
what is predicted, especially in the $\Rey=420$ case, shown 
in \Fig{fig:calcEta_k1.5_421}. Here, \Eq{eq:expectedCapEff} 
cannot predict the somewhat random jumps in impaction 
efficiency observed at the lowest Stokes numbers. 
The assumption that the impaction efficiency in the inertial 
impaction mode is not affected by the turbulence, seems to 
be reasonable. 

Concluding from the results depicted in 
\Figs{fig:calcEta_k1.5_421}{fig:calcEta_k1.5_1685}, the 
variances of the effective Stokes numbers explain the increased 
impaction efficiencies in the boundary stopping mode. 
The curves of calculated values $E(\eta(\St_{\mathrm{eff}}))$ 
(shown by dashed-dotted lines in the figures)
follow the turbulent simulation curves in both figures. 

With $\kf=5$ and $\kf=15$ at $\Rey=420$ 
(\Figs{fig:calcEta_k5_421}{fig:calcEta_k15_421}, respectively), 
the capture efficiency in the turbulent cases does not 
deviate much from the laminar. Therefore, the corresponding scenarios at $\Rey=1685$ 
(\Figs{fig:calcEta_k5_1685}{fig:calcEta_k15_1685}, respectively)
are studied when discussing the validity of \Eq{eq:expectedCapEff}
for $\kf=5$ and $\kf=15$. 

As seen in \Fig{fig:calcEta_k5_1685}, the calculated expected 
capture efficiency does not match perfectly with the simulated values for $\kf=5$. In the lower region of the boundary stopping mode, the expected capture efficiency is found to be above the 
values from the turbulent simulation. In the range 
$0.2 \lesssim \St \lesssim 0.25$, the calculated values match
with the simulation values, while for 
$0.25 \lesssim \St \lesssim 0.90$, the simulation 
values exceed the calculated values.



Turning to \Fig{fig:calcEta_k15_1685}, the results from the
simulation are seen to deviate considerably from the expected values
also in the boundary stopping mode, up to $\St\approx 0.22$; the
calculated $E(\eta(\St_{\mathrm{eff}}))$ is too high. For Stokes numbers up
to $\St\approx 0.15$
both the laminar results and the results calculated from \Eq{eq:expectedCapEff}
are found to be higher than the simulation results.

The unpredictable nature of the results seen only in the boundary
interception mode ($\St \le 0.15$) for the $\kf=1.5$ seems to 
become more present also
in the boundary stopping mode as $\kf$ is raised. This indicates that
the behaviour of the turbulent eddies close to the boundary layer of
the cylinder has a larger effect on particles of increasing Stokes
numbers as the forcing scale is decreased.  


\subsection{Particle Clustering}
\label{clustering}
\begin{figure}[h!]
  \begin{center}
    \subfloat[$\kf=15$]{\includegraphics[width=0.4\textwidth]{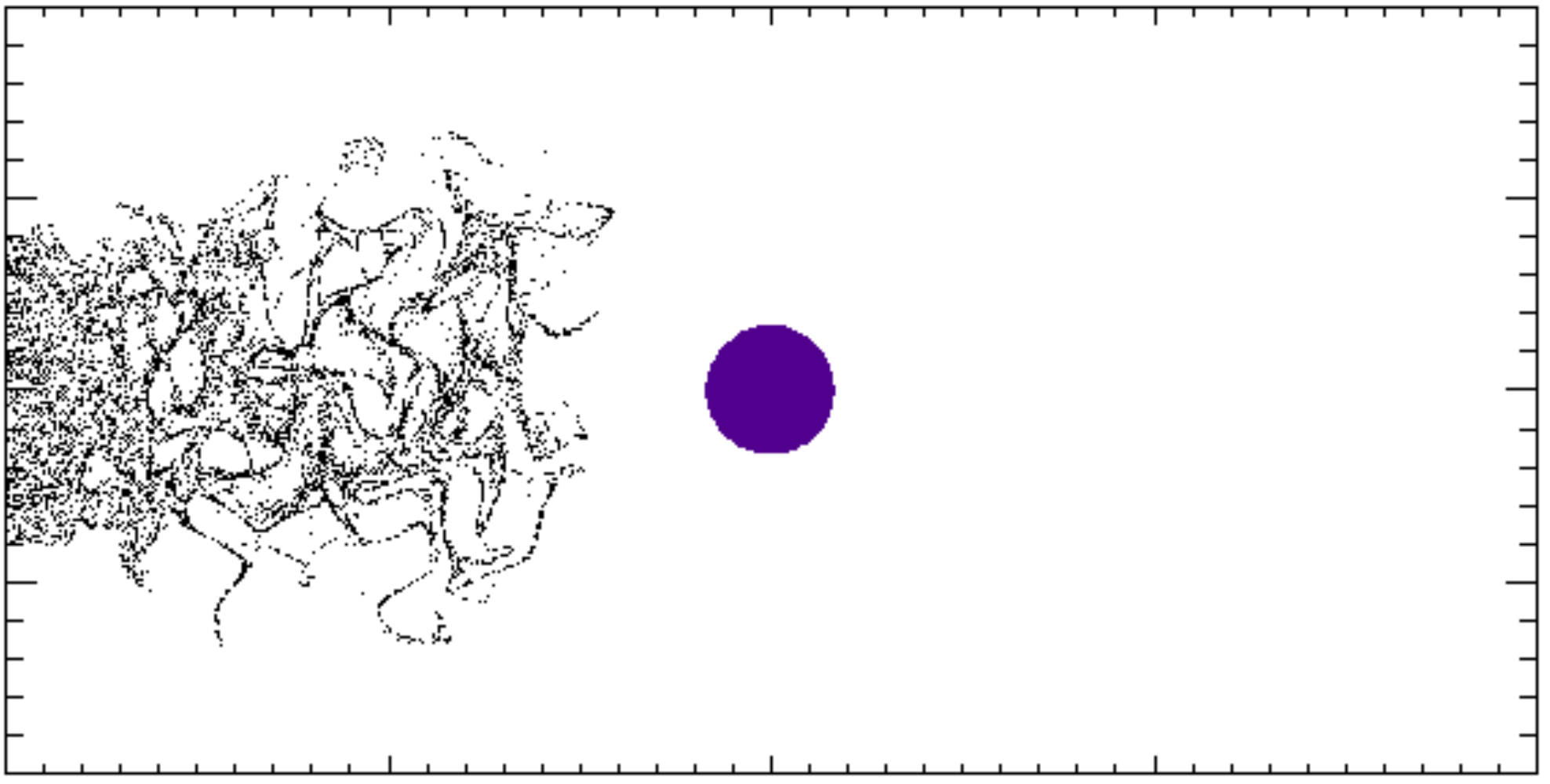}}\\
    \subfloat[$\kf=5$]{\includegraphics[width=0.4\textwidth]{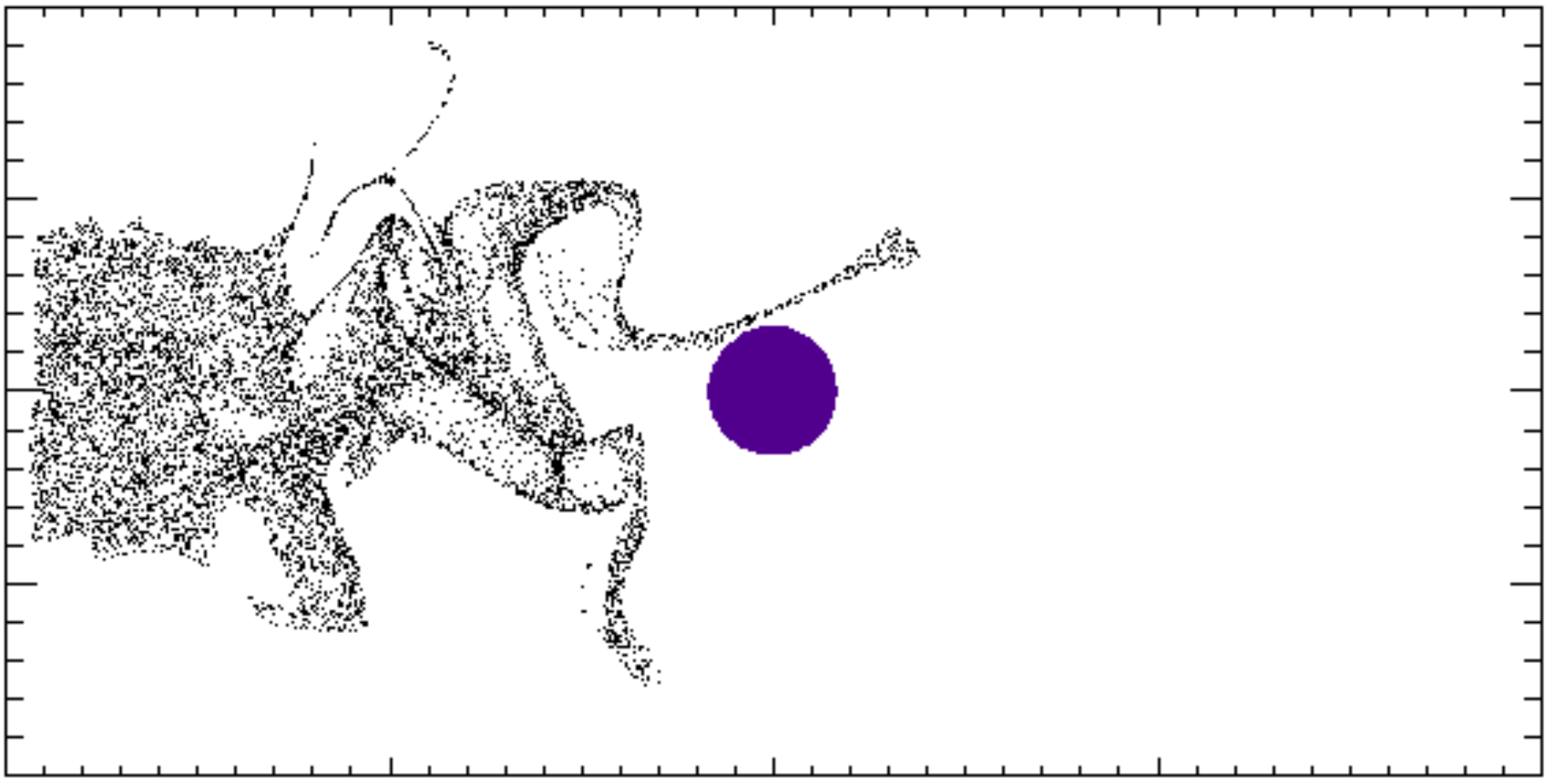}}\\
    \subfloat[$\kf=1.5$]{\includegraphics[width=0.4\textwidth]{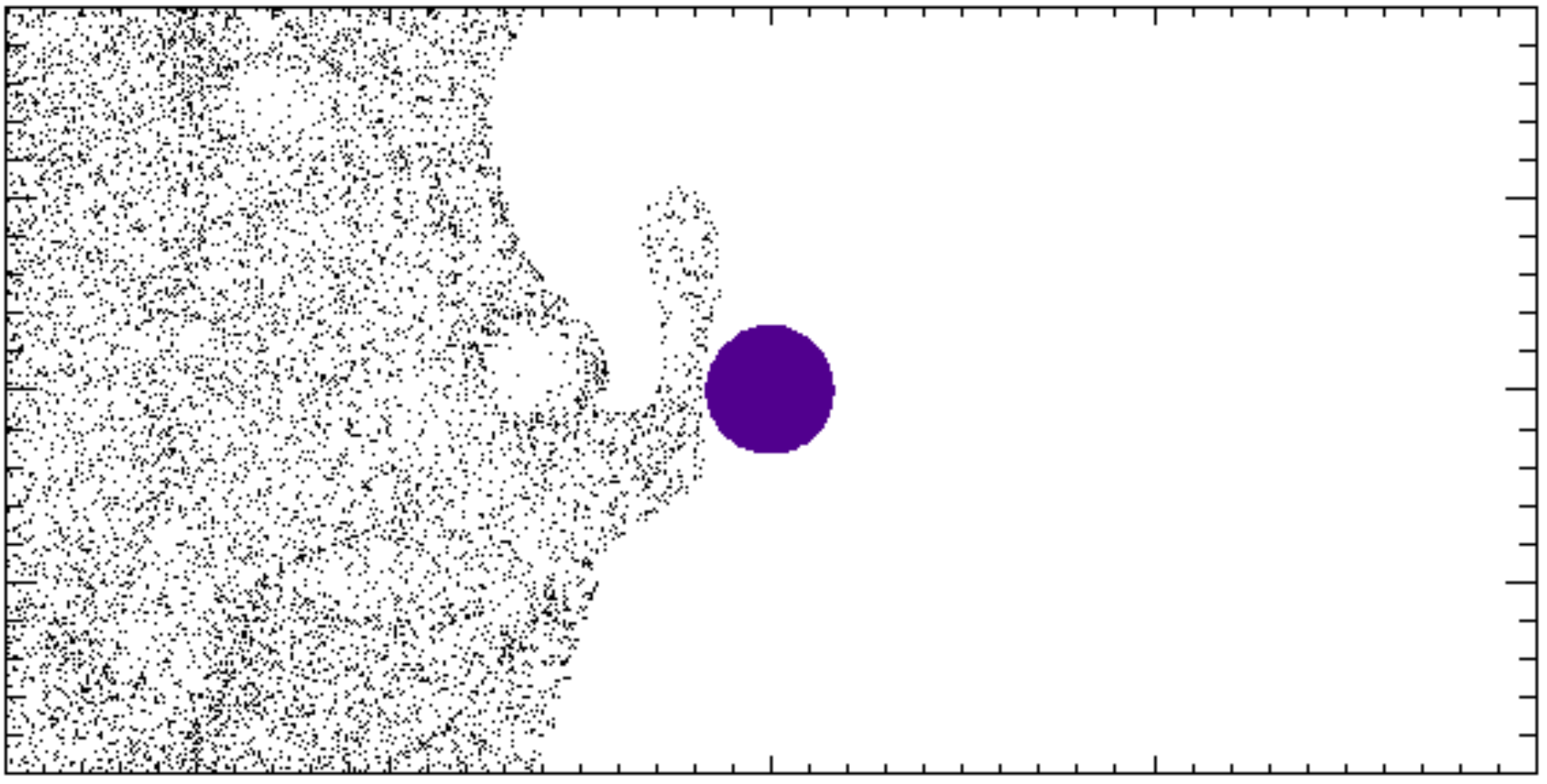}}
  \end{center}
  \caption{Clustering of particles with $\St=0.30$, from turbulent cases run at $\Rey=1685$.}
 \label{fig:clustering1}
\end{figure}

The mechanism behind particle clustering,
or preferential concentration of particles, 
can be explained as follows. The same explanation 
applies also for backside impaction, as described in the 
next section. Since the forcing scale 
$l_{\mathrm{f}}=L_{\mathrm{box}}/\kf$ is a characteristic 
length for the turbulence, the characteristic eddy time 
can be defined as
\begin{equation}
\tau_{\mathrm{eddy}}=\frac{l_{\mathrm{f}}}{u_{\mathrm{rms}}}=
\frac{L_{\mathrm{box}}}{u_{\mathrm{rms}}\kf}. 
\label{eq:tauEddy}
\end{equation}

The product $u_{\mathrm{rms}}\kf$, values of which 
are found in table \ref{tab:urms}, determines 
$\tau_{\mathrm{eddy}}$ for the different turbulent cases.

For $\Rey=1685$, the particles exhibited clustering. The 
$\Rey=420$ cases are 
not considered, as the phenomenon is less apparent 
here due to the stronger damping of the turbulent eddies. 
The eddy Stokes number can be introduced as 
\begin{equation}
\label{eq:eddyStokes}
\St_{\mathrm{eddy}}=\frac{\tau_{\mathrm{p}}}{\tau_{\mathrm{eddy}}}
=\frac{\tau_{\mathrm{f}}}{\tau_{\mathrm{eddy}}}\St.
\end{equation}
Here, the definition 
$\St=\tau_{\mathrm{p}}/\tau_{\mathrm{f}}$ has been used. 
With the characteristic fluid time $\tau_{\mathrm{f}}=D/U_0$ 
and the characteristic eddy time $\tau_{\mathrm{eddy}}$ 
given by Eq. (\ref{eq:tauEddy}), Eq. (\ref{eq:eddyStokes}) 
can be expressed as 
\begin{equation}
\label{eq:eddyStokes2}
\St_{\mathrm{eddy}}=\frac{D}{U_0}\frac{u_{\mathrm{rms}}\kf}
{L_{\mathrm{box}}}\St.
\end{equation}
It is known that $\tau_{\mathrm{p}}\sim\tau_{\mathrm{eddy}}$, i.e. 
$\St_{\mathrm{eddy}}\sim 1$, is needed for
particle clustering to take place. 
This is because if $\tau_{\mathrm{eddy}} \gg \tau_{\mathrm{p}}$, the eddy 
turn-over time is too slow for the centrifugal 'force' to 
throw the particle towards the wall. 
When $\tau_{\mathrm{eddy}} \ll \tau_{\mathrm{p}}$, which is 
the case for large particles, the particle will not have 
time to respond to the fast turbulent motions and thus will not 
obtain the acceleration needed. 
As it is
$u_{\mathrm{rms}}\kf$ in the prefactor in front of $\St$ in Eq.
(\ref{eq:eddyStokes2}) that is varying between the different turbulent
cases, this product determines for which Stokes numbers clustering
will be large, namely for those implying $\St_{\mathrm{eddy}}\sim
1$. For the Stokes numbers considered in this work, this implies that
particle clustering is largest at $\kf=15$, while the effect gets
smaller for decreasing $\kf$. This can be seen in the last column 
of \Tab{tab:urms} which shows $\St_{\rm eddy}$. 
The discussion of the role of the
vorticity in the next section is related to this explanation, since it
is the vorticity of the turbulent eddies that gives rise to
$\St_{\mathrm{eddy}}$, with the magnitude of vorticity $\omega\propto
u_{\mathrm{rms}}\kf$.

\begin{table}[h!]
\caption{Some important parameters for the turbulence simulations. Note that
the eddy Stokes number in the last column is for particles with Stokes number
0.3.}
\label{tab:urms}
\begin{center}
\begin{tabular}{c|c|c|c|c}
&$\kf$&$u_{\mathrm{rms}}$&$u_{\mathrm{rms}}\kf$&$\St_{\rm eddy}(\St=0.3)$\\
\hline
\multirow{3}{*}{$\Rey=420$} & 15 & 0.17 & 2.59 & 0.13\\
& 5   & 0.27 & 1.34 &0.07\\
& 1.5 & 0.42 & 0.64 &0.03\\
\hline
\multirow{3}{*}{$\Rey=1685$} & 15 & 0.24 & 3.57 & 0.18\\
& 5   & 0.28 & 1.40 & 0.07\\
& 1.5 & 0.29 & 0.44 & 0.02\\
\hline
\end{tabular}
\end{center}
\end{table}

\subsection{Backside Impaction}

\begin{figure}[h!]
\centering
\includegraphics[width=0.4\textwidth]{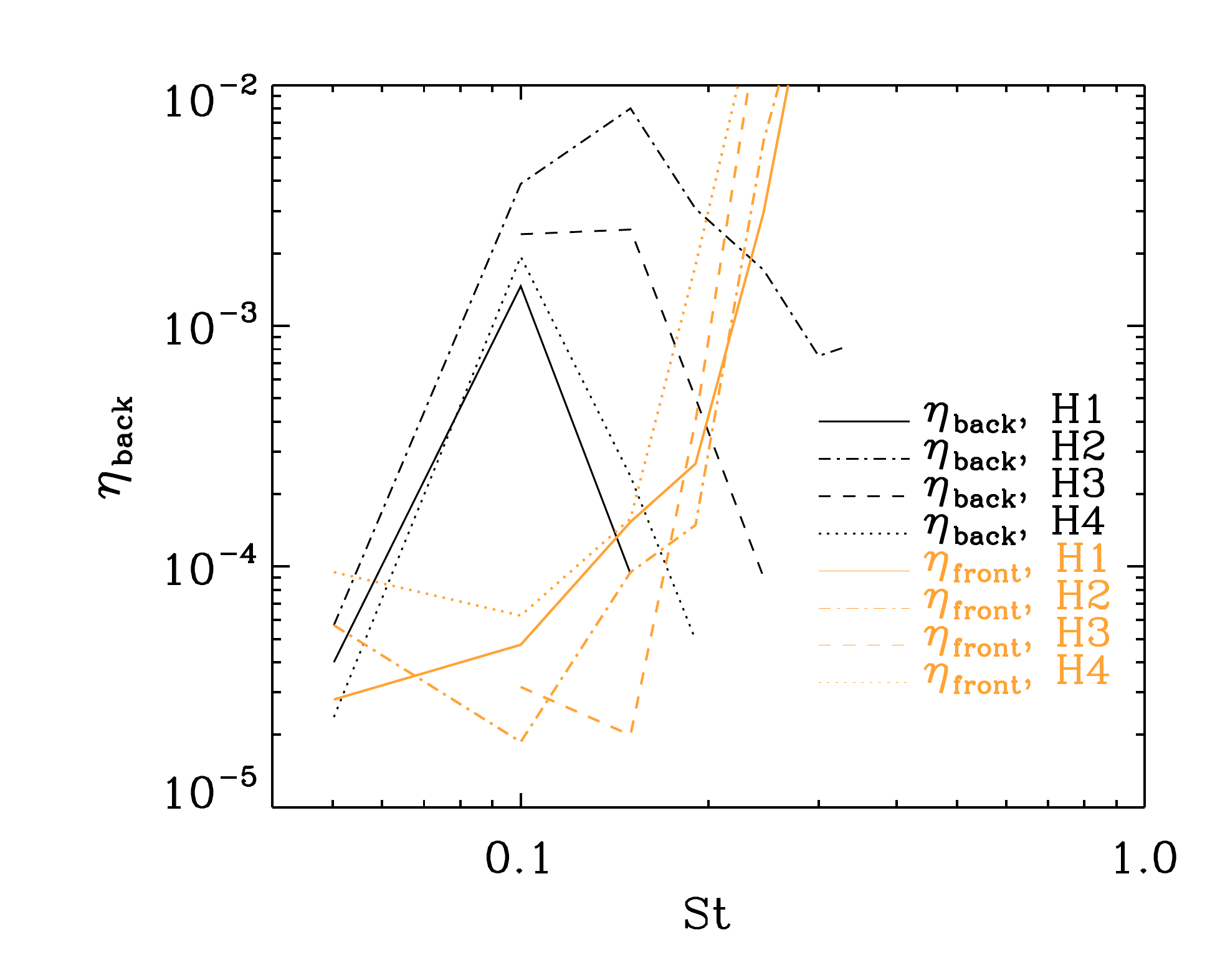}
\caption{Back side impaction efficiency for $\Rey=1685$.}
\label{fig:calcEff_back_Re1685}
\end{figure}

To further investigate the effects of turbulent eddies 
on the particles, the impaction on the backside of the cylinder, 
$\eta_{\mathrm{back}}$, 
is plotted in \Fig{fig:calcEff_back_Re1685}.
In the $\Rey=420$ simulations, backside impaction takes place 
only to a small extent and is therefore not shown here.
For $\Rey=1685$, more general trends in the backside 
impaction are seen. In all cases, the backside impaction 
is larger than the front-side impaction for 
$\St\lesssim 0.14$, and for the turbulence with the smallest integral 
scale $\eta_{\rm back}>\eta_{\rm front}$ for $\St\lesssim 0.23$. The general 
behaviour for the $\Rey=1685$ cases is that $\capback$ 
increases with increasing Stokes numbers from its initial minimum until 
it reaches 
the maximum which is in the range $0.1<\St<0.15$, from where it decays.

For a particle to impact on the backside of the cylinder it must first
be captured by an eddy bringing it to the back side of the cylinder. This
could be either a typical von K\'{a}rm\'{a}n eddy or a large scale turbulent eddy.
Being in the wake on the back side of the cylinder a particle will typically
impact due to an effect very similar to turbophoresis, which is encountered
in turbulent channel flows.
Which particles are impacting on the backside is, in a similar manner as for
particle clustering (\Sec{clustering}), 
dependent on the eddy Stokes number.

In the laminar cases, $\tau_{\mathrm{eddy}}$ is given by 
the dimensions of the cylinder, as backside impaction in 
the laminar cases is caused only by the rotational motion 
of the von K\'{a}rm\'{a}n eddies in the wake. These wake vortices also play a 
role in the turbulent $\capback$, but the 
differences in the turbulent $\capback$ can be explained 
by \Eq{eq:eddyStokes2}. 

\section{Conclusion}
Direct Numerical Simulations were run to study the influence 
of isotropic turbulence on the impaction 
of particles on a cylinder in a cross flow at 
two different Reynolds numbers, $\Rey=420$ and $\Rey=1685$.
The turbulence was simulated on 
a three-dimensional domain, with forcing at three different wave numbers, 
i.e. at varying integral scales. 
After reaching a homogeneous steady state, the turbulence was 
imposed on the main flow domain, with one-way 
coupling between the fluid and the particles.
It was checked that for $\Rey=420$ a two dimensional main 
flow domain gave qualitatively the
same results as a three dimensional main flow domain. 

The impaction efficiency of 
particles on the cylinder front side was seen to be greater in 
the turbulent cases, compared to laminar reference cases in 
the boundary stopping mode. This was found to be related to 
the statistical variance of $\St_{\mathrm{eff}}$, where the effective Stokes number takes into account the fluctuating particle velocities due to turbulent velocity variations. The laminar front side impaction efficiency rapidly increases with increasing Stokes numbers in the boundary stopping mode, which yields a larger impaction efficiency when turbulence is introduced. The 
relation to the variance of $\St_{\mathrm{eff}}$ was clearest for the 
largest integral scale, and was less present for 
decreasing integral scales. 

As the Reynolds number 
was increased from $\Rey=420$ to $\Rey=1685$, the impact of 
the turbulence on the particle impaction became 
more dominant. For the smallest particles, in the 
boundary interception mode, the turbulence also lead to 
changes in the front side impaction efficiency. However, these 
changes could not be explained by the fluctuations in the 
effective Stokes number. This indicates that the turbulence 
causes some not fully understood mechanism influencing the motion 
of the smallest particles in the close neighborhood of 
the boundary layer of the cylinder. 

Particle impaction efficiency on the backside of the cylinder in the 
$\Rey=1685$ turbulent cases was seen to be relatively large. 
This was particularly true for the turbulence with the 
smaller integral scale, which, due to its strong vorticity, 
had the largest backside impaction efficiency. 

Since the turbulent eddies in the $\Rey=1685$ cases were 
well sustained throughout the two-dimensional domain, 
the particles dispersed in the turbulent flow exhibited 
clustering, or preferential concentration. Particle clustering 
is caused by the same mechanisms as backside impaction. 
Thus, the magnitude of vorticity, and slow dissipation 
thereof, is of crucial importance for the clustering to 
come into play.

\begin{acknowledgements}
This work was supported by the Norwegian Research Council project 
PAFFrx (186933) and by the competence building project KRAV 
(Enabling small-scale biomass CHP in Norway) funded by the Research 
Council of Norway, five Norwegian Industry Partners and SINTEF Energy Research.\\
The authors acknowledge the invaluable discussions with Steinar Kragset.
\end{acknowledgements}

\end{document}